\documentclass[12pt]{article}
\usepackage{amssymb}
\renewcommand{\theequation}{\arabic{section}.\arabic{equation}}

\begin{document}

%************************** Text Begins here ******************************

%  Greek letters

\def\a{\alpha}
\def\b{\beta}
\def\d{\delta}
\def\e{\epsilon}
\def\g{\gamma}
\def\h{\mathfrak{h}}
\def\k{\kappa}
\def\l{\lambda}
\def\o{\omega}
\def\p{\wp}
\def\r{\rho}
\def\t{\tau}
\def\s{\sigma}
\def\z{\zeta}
\def\x{\xi}
 \def\A{{\cal{A}}}
 \def\B{{\cal{B}}}
 \def\C{{\cal{C}}}
 \def\D{{\cal{D}}}
\def\G{\Gamma}
\def\K{{\cal{K}}}
\def\O{\Omega}
\def\R{\bar{R}}
\def\T{{\cal{T}}}
\def\L{\Lambda}
\def\f{E_{\tau,\eta}(sl_2)}
\def\E{E_{\tau,\eta}(sl_n)}
\def\Zb{\mathbb{Z}}
\def\Cb{\mathbb{C}}

\def\R{\overline{R}}
% Shorthands for \begin{equation} and the like

\def\beq{\begin{equation}}
\def\eeq{\end{equation}}
\def\bea{\begin{eqnarray}}
\def\eea{\end{eqnarray}}
\def\ba{\begin{array}}
\def\ea{\end{array}}
\def\no{\nonumber}
\def\le{\langle}
\def\re{\rangle}
\def\lt{\left}
\def\rt{\right}

\newtheorem{Theorem}{Theorem}
\newtheorem{Definition}{Definition}
\newtheorem{Proposition}{Proposition}
\newtheorem{Lemma}{Lemma}
\newtheorem{Corollary}{Corollary}
\newcommand{\proof}[1]{{\bf Proof. }
        #1\begin{flushright}$\Box$\end{flushright}}

\baselineskip=20pt

%%%%%%%%%%%%%%%%%%%%%%%%%%%%%%%%%%%%%%%%%%%%%%%%%%%%%%%%%%%%
%                                                          %
%  Title page                                              %
%                                                          %
%%%%%%%%%%%%%%%%%%%%%%%%%%%%%%%%%%%%%%%%%%%%%%%%%%%%%%%%%%%%
\newfont{\elevenmib}{cmmib10 scaled\magstep1}
\newcommand{\preprint}{
   \begin{flushleft}
     %\elevenmib Yukawa\, Institute\, Kyoto\\
   \end{flushleft}\vspace{-1.3cm}
   \begin{flushright}\normalsize
  % \sf  YITP-03-53\\
  %  {\tt hep-th/0703222} \\ March 2007
   \end{flushright}}
\newcommand{\Title}[1]{{\baselineskip=26pt
   \begin{center} \Large \bf #1 \\ \ \\ \end{center}}}
\newcommand{\Author}{\begin{center}
   \large \bf
Wen-Li Yang${}^{a}$, ~Xi Chen${}^{a}$, ~Jun Feng${}^{a}$,~Kun Hao${}^{a}$,~Bo-Yu Hou${}^{a}$,\\~Kang-Jie Shi ${}^a$
 ~and~Yao-Zhong Zhang ${}^b$
 \end{center}}
\newcommand{\Address}{\begin{center}

     ${}^a$ Institute of Modern Physics, Northwest University,
     Xian 710069, P.R. China\\
     ${}^b$ The University of Queensland, School of Mathematics and Physics,  Brisbane, QLD 4072,
     Australia\\
    E-mails: wlyang@nwu.edu.cn, \,\,chenxi0905@yahoo.cn, \,\,grammophon@163.com, \,\,hoke72@163.com, \,\,byhou@nwu.edu.cn,
    \,\,kjshi@nwu.edu.cn \,\,and\,\,yzz@maths.uq.edu.au
   \end{center}}
\newcommand{\Accepted}[1]{\begin{center}
   {\large \sf #1}\\ \vspace{1mm}{\small \sf Accepted for Publication}
   \end{center}}

\preprint
\thispagestyle{empty}
\bigskip\bigskip\bigskip

\Title{Determinant representations of scalar products for the open
XXZ chain with non-diagonal boundary terms } \Author

\Address
\vspace{1cm}

\begin{abstract}
With the help of the F-basis provided by the Drinfeld twist or factorizing
F-matrix for the open XXZ spin chain with non-diagonal boundary
terms, we obtain the determinant representations of the scalar
products of Bethe states of the model.

\vspace{1truecm} \noindent {\it PACS:} 03.65.Fd; 04.20.Jb;
05.30.-d; 75.10.Jm

\end{abstract}
\newpage
%%%%%%%%%%%%%%%%%%%%%%%%%%%%%%%%%%%%%%%%%%%%%%%%%%%%%%%%%%%%%%%
%                                                             %
%  1. Introduction                                            %
%                                                             %
%%%%%%%%%%%%%%%%%%%%%%%%%%%%%%%%%%%%%%%%%%%%%%%%%%%%%%%%%%%%%%%
\section{Introduction}
\label{intro} \setcounter{equation}{0}
The computation of correlation functions (or scalar products of Bethe states)
is one of major challenging problems in the theory of quantum integrable models
\cite{Smi92, Kor93}. There are two approaches in the literature for computing the correlation
functions of a quantum integrable model. One is the vertex operator method (see e.g.
\cite{Fre92,Dav93,Koy94,Hou97,Yan99,Hou99}) which works only on an infinite lattice,
and another one is based on the detailed analysis of the structure of the Bethe states
\cite{Kor82, Ize87}. As for the second approach which usually works for models with
finite size, it is well known that
in the framework of quantum inverse scattering method (QISM) \cite{Kor93} Bethe states are
obtained by applying pseudo-particle creation operators to
reference state (pseudo-vacuum). However, the apparently simple action of creation operators is
plagued with non-local effects arising from polarization clouds or
compensating exchange terms on the level of local operators. This makes the direct
calculation of correlation functions of models with finite size challenging.

Progress has recently
been made on the second approach with the help of the Drinfeld twists or
factorizing F-matrices \cite{Dri83}. Working in the F-basis provided by the F-matrices,
the authors in \cite{Mai00, Kit99} managed to calculate the form factors and correlation
functions of the XXX and XXZ chains  with periodic boundary condition (or closed chains)
analytically and expressed them in determinant forms. Then the determinant representation
of the scalar products and correlation functions of the supersymmetric t-J model \cite{Zha06}
and its q-deformed model \cite{Zha06-1} with periodic boundary condition was obtained within the
corresponding F-basis given in \cite{Yan04-2}.

It was  noticed  \cite{Wan02,Kit07} that the F-matrices of
the closed XXX and XXZ chains also  make the pseudo-particle
creation operators of the open XXX and XXZ chains with diagonal
boundary terms polarization free. This is mainly due to the fact
that the closed chain and the corresponding open chain with
diagonal boundary terms share the same reference state
\cite{Skl88}. However, the story for the open XXZ chain with
non-diagonal boundary terms is quite different
\cite{Nep04,Cao03,Yan04,Gal05,Gie05,Gie05-1,Yan04-1,Baj06,Yan05,Doi06,Mur06,Bas07,Gal08,Mur09}.
Firstly, the reference state (all spin up state) of the closed
chain is no longer a reference state of the open chain with
non-diagonal boundary terms \cite{Cao03,Yan04,Yan04-1}. Secondly,
at least two reference states (and thus two sets of Bethe states) are
needed \cite{Yan07} for the open XXZ chain with non-diagonal
boundary terms in order to obtain its complete spectrum
\cite{Nep03,Yan06}. As a consequence, the F-matrix found in
\cite{Mai00} is no longer the {\it desirable} F-matrix for the
open XXZ chain with non-diagonal boundary terms.

Very recently, we have succeeded in obtaining the factorizing
F-matrices for the open XXZ chain with integrable boundary
conditions given  by the non-diagonal K-matrices
(\ref{K-matrix-2-1}) and (\ref{K-matrix-6}) \cite{Yan09}.  In this paper,
we shall investigate  the determinant representations of the scalar products
of the Bethe states of the open XXZ chain with non-diagonal boundary terms
with the help of the associated F-matrices.

The paper is organized as follows.  In section 2, we briefly
describe the open XXZ chain with non-diagonal boundary terms, and
introduce the pseudo-particle creation operators and the two sets
of Bethe states of the model. In section 3, we introduce the face picture
of the model and express the scalar products in terms of the operators in the
face picture. In section 4, we present the F-matrix of the
open XXZ chain in the face picture and give  the completely
symmetric and polarization free representations of the
pseudo-particle creation/annihilation operators  in the F-basis. In section 5,
with the help of the F-basis, we obtain the determinant representations of the scalar
products of Bethe states. In
section 6, we summarize our results and give some discussions.

%%%%%%%%%%%%%%%%%%%%%%%%%%%%%%%%%%%%%%%%%%%%%%%%%%%%%%%%%%%%%%%
%                                                             %
%  2. The inhomogeneous spin-$\frac{1}{2}$                    %
%                    XXZ open chain                           %
%                                                             %
%                                                             %
%%%%%%%%%%%%%%%%%%%%%%%%%%%%%%%%%%%%%%%%%%%%%%%%%%%%%%%%%%%%%%%

\section{ The inhomogeneous spin-$\frac{1}{2}$ XXZ open chain}
\label{XXZ} \setcounter{equation}{0}

Throughout, $V$ denotes a two-dimensional linear space. The
spin-$\frac{1}{2}$ XXZ chain can be constructed from the
well-known six-vertex model R-matrix $\R(u)\in {\rm End}(V\otimes
V)$ \cite{Kor93} given by \bea
\R(u)=\lt(\begin{array}{llll}1&&&\\&b(u)&c(u)&\\
&c(u)&b(u)&\\&&&1\end{array}\rt).\label{r-matrix}\eea The
coefficient functions read: $b(u)=\frac{\sin u}{\sin(u+\eta)}$,
$c(u)=\frac{\sin\eta}{\sin(u+\eta)}$. Here we assume  $\eta$ is a
generic complex number. The R-matrix satisfies the quantum
Yang-Baxter equation (QYBE), \bea
R_{1,2}(u_1-u_2)R_{1,3}(u_1-u_3)R_{2,3}(u_2-u_3)
=R_{2,3}(u_2-u_3)R_{1,3}(u_1-u_3)R_{1,2}(u_1-u_2),\label{QYB}\eea
and the unitarity, crossing-unitarity and quasi-classical
properties \cite{Yan04-1}. We adopt the standard notations: for
any matrix $A\in {\rm End}(V)$ , $A_j$ (or $A^j$) is an embedding
operator in the tensor space $V\otimes V\otimes\cdots$, which acts
as $A$ on the $j$-th space and as identity on the other factor
spaces; $R_{i,j}(u)$ is an embedding operator of R-matrix in the
tensor space, which acts as identity on the factor spaces except
for the $i$-th and $j$-th ones.

One introduces the ``row-to-row"  (or one-row ) monodromy matrix
$T(u)$, which is an $2\times 2$ matrix with elements being
operators acting on $V^{\otimes N}$, where $N=2M$ ($M$ being a
positive integer),\bea
T_0(u)=\R_{0,N}(u-z_N)\R_{0,N-1}(u-z_{N-1})\cdots
\R_{0,1}(u-z_1).\label{Mon-V}\eea Here $\{z_j|j=1,\cdots,N\}$ are
arbitrary free complex parameters which are usually called
inhomogeneous parameters.

Integrable open chain can be constructed as follows \cite{Skl88}.
Let us introduce a pair of K-matrices $K^-(u)$ and $K^+(u)$. The
former satisfies the reflection equation (RE)
 \bea &&\R_{1,2}(u_1-u_2)K^-_1(u_1)\R_{2,1}(u_1+u_2)K^-_2(u_2)\no\\
 &&~~~~~~=
K^-_2(u_2)\R_{1,2}(u_1+u_2)K^-_1(u_1)\R_{2,1}(u_1-u_2),\label{RE-V}\eea
and the latter  satisfies the dual RE \bea
&&\R_{1,2}(u_2-u_1)K^+_1(u_1)\R_{2,1}(-u_1-u_2-2\eta)K^+_2(u_2)\no\\
&&~~~~~~=
K^+_2(u_2)\R_{1,2}(-u_1-u_2-2\eta)K^+_1(u_1)\R_{2,1}(u_2-u_1).
\label{DRE-V}\eea For open spin-chains, instead of the standard
``row-to-row" monodromy matrix $T(u)$ (\ref{Mon-V}), one needs to
consider  the
 ``double-row" monodromy matrix $\mathbb{T}(u)$
\bea
  \mathbb{T}(u)=T(u)K^-(u)\hat{T}(u),\quad \hat{T}(u)=T^{-1}(-u).
  \label{Mon-V-0}
\eea Then the double-row transfer matrix of the XXZ chain with
open boundary (or the open XXZ chain) is given by \bea
\t(u)=tr(K^+(u)\mathbb{T}(u)).\label{trans}\eea The QYBE and
(dual) REs lead to that the transfer matrices with different
spectral parameters commute with each other \cite{Skl88}:
$[\t(u),\t(v)]=0$. This ensures the integrability of the open XXZ
chain.

In this paper, we will consider the K-matrix $K^{-}(u)$ which is a
generic solution to the RE (\ref{RE-V}) associated the six-vertex
model R-matrix  \cite{Veg93,Gho94}
\bea K^-(u)=\lt(\begin{array}{ll}k_1^1(u)&k^1_2(u)\\
k^2_1(u)&k^2_2(u)\end{array}\rt)\equiv K(u).\label{K-matrix}\eea
The coefficient functions are \bea && k^1_1(u)=
\frac{\cos(\l_1-\l_2) -\cos(\l_1+\l_2+2\xi)e^{-2iu}}
{2\sin(\l_1+\xi+u)
\sin(\l_2+\xi+u)},\no\\
&&k^1_2(u)=\frac{-i\sin(2u)e^{-i(\l_1+\l_2)} e^{-iu}}
{2\sin(\l_1+\xi+u) \sin(\l_2+\xi+u)},\no\\
&&k^2_1(u)=\frac{i\sin(2u)e^{i(\l_1+\l_2)} e^{-iu}}
{2\sin(\l_1+\xi+u) \sin(\l_2+\xi+u)}, \no\\
&& k^2_2(u)=\frac{\cos(\l_1-\l_2)e^{-2iu}- \cos(\l_1+\l_2+2\xi)}
{2\sin(\l_1+\xi+u)\sin(\l_2+\xi+u)}.\label{K-matrix-2-1} \eea At
the same time, we introduce  the corresponding {\it dual\/}
K-matrix $K^+(u)$ which is a generic solution to the dual
reflection equation (\ref{DRE-V}) with a particular choice of the
free boundary parameters:
\bea K^+(u)=\lt(\begin{array}{ll}{k^+}_1^1(u)&{k^+}^1_2(u)\\
{k^+}^2_1(u)&{k^+}^2_2(u)\end{array}\rt)\label{DK-matrix}\eea with
the matrix elements \bea && {k^+}^1_1(u)=
\frac{\cos(\l_1-\l_2)e^{-i\eta}
-\cos(\l_1+\l_2+2\bar{\xi})e^{2iu+i\eta}}
{2\sin(\l_1+\bar{\xi}-u-\eta)
\sin(\l_2+\bar{\xi}-u-\eta)},\no\\
&&{k^+}^1_2(u)=\frac{i\sin(2u+2\eta)e^{-i(\l_1+\l_2)}
e^{iu-i\eta}} {2\sin(\l_1+\bar{\xi}-u-\eta)
\sin(\l_2+\bar{\xi}-u-\eta)},
\no\\
&&{k^+}^2_1(u)=\frac{-i\sin(2u+2\eta)e^{i(\l_1+\l_2)}
e^{iu+i\eta}}
{2\sin(\l_1+\bar{\xi}-u-\eta) \sin(\l_2+\bar{\xi}-u-\eta)}, \no\\
&& {k^+}^2_2(u)=\frac{\cos(\l_1-\l_2)e^{2iu+i\eta}-
\cos(\l_1+\l_2+2\bar{\xi})e^{-i\eta}}
{2\sin(\l_1+\bar{\xi}-u-\eta)\sin(\l_2+\bar{\xi}-u-\eta)}.\label{K-matrix-6}
\eea The K-matrices depend on four free boundary parameters
$\{\l_1,\,\l_2,\,\xi,\,\bar{\xi}\}$. It is
very convenient to introduce a vector $\l\in V$ associated with
the boundary parameters $\{\l_i\}$, \bea
 \l=\sum_{k=1}^2\l_k\e_k, \label{boundary-vector}
\eea  where $\{\e_i,\,i=1,2\}$ form the orthonormal basis of $V$
such that $\langle \e_i,\e_j\rangle=\d_{ij}$.

%%%%%%%%%%%%%%%%%%%%%%%%%%%%%%%%%%%%%%%%%%%%%%%%%%%%%%%%%%%%%%%%%%%%%%%%%%%%%%%%

\subsection{Vertex-face correspondence}

Let us briefly review the face-type R-matrix associated with the
six-vertex model.

Set \bea \hat{\imath}=\e_i-\overline{\e},~~\overline{\e}=
\frac{1}{2}\sum_{k=1}^{2}\e_k, \quad i=1,2,\qquad {\rm then}\,
\sum_{i=1}^2\hat{\imath}=0. \label{fundmental-vector} \eea Let
$\h$ be the Cartan subalgebra of $A_{1}$ and $\h^{*}$ be its dual.
A finite dimensional diagonalizable  $\h$-module is a complex
finite dimensional vector space $W$ with a weight decomposition
$W=\oplus_{\mu\in \h^*}W[\mu]$, so that $\h$ acts on $W[\mu]$ by
$x\,v=\mu(x)\,v$, $(x\in \h,\,v\in\,W[\mu])$. For example, the
non-zero weight spaces of the fundamental representation
$V_{\L_1}=\Cb^2=V$ are
\bea
 W[\hat{\imath}]=\Cb \e_i,~i=1,2.\label{Weight}
\eea

For a generic $m\in V$, define \bea m_i=\langle m,\e_i\rangle,
~~m_{ij}=m_i-m_j=\langle m,\e_i-\e_j\rangle,~~i,j=1,2.
\label{Def1}\eea Let $R(u,m)\in {\rm End}(V\otimes V)$ be the
R-matrix of the six-vertex SOS model, which is trigonometric limit
of the eight-vertex SOS model \cite{Bax82} given by
\bea
R(u;m)\hspace{-0.1cm}=\hspace{-0.1cm}
\sum_{i=1}^{2}R(u;m)^{ii}_{ii}E_{ii}\hspace{-0.1cm}\otimes\hspace{-0.1cm}
E_{ii}\hspace{-0.1cm}+\hspace{-0.1cm}\sum_{i\ne
j}^2\lt\{R(u;m)^{ij}_{ij}E_{ii}\hspace{-0.1cm}\otimes\hspace{-0.1cm}
E_{jj}\hspace{-0.1cm}+\hspace{-0.1cm}
R(u;m)^{ji}_{ij}E_{ji}\hspace{-0.1cm}\otimes\hspace{-0.1cm}
E_{ij}\rt\}, \label{R-matrix} \eea where $E_{ij}$ is the matrix
with elements $(E_{ij})^l_k=\d_{jk}\d_{il}$. The coefficient
functions are \bea
 &&R(u;m)^{ii}_{ii}=1,~~
   R(u;m)^{ij}_{ij}=\frac{\sin u\sin(m_{ij}-\eta)}
   {\sin(u+\eta)\sin(m_{ij})},~~i\neq j,\label{Elements1}\\
 && R(u;m)^{ji}_{ij}=\frac{\sin\eta\sin(u+m_{ij})}
    {\sin(u+\eta)\sin(m_{ij})},~~i\neq j,\label{Elements2}
\eea  and $m_{ij}$ is defined in (\ref{Def1}). The R-matrix
satisfies the dynamical (modified) quantum Yang-Baxter equation
(or the star-triangle relation) \cite{Bax82}
\begin{eqnarray}
&&R_{1,2}(u_1-u_2;m-\eta h^{(3)})R_{1,3}(u_1-u_3;m)
R_{2,3}(u_2-u_3;m-\eta h^{(1)})\no\\
&&\qquad =R_{2,3}(u_2-u_3;m)R_{1,3}(u_1-u_3;m-\eta
h^{(2)})R_{1,2}(u_1-u_2;m).\label{MYBE}
\end{eqnarray}
Here we have adopted
\bea R_{1,2}(u,m-\eta h^{(3)})\,v_1\otimes
v_2 \otimes v_3=\lt(R(u,m-\eta\mu)\otimes {\rm id }\rt)v_1\otimes
v_2 \otimes v_3,\quad {\rm if}\, v_3\in W[\mu]. \label{Action}
\eea Moreover, one may check that the R-matrix satisfies  weight
conservation condition, \bea
  \lt[h^{(1)}+h^{(2)},\,R_{1,2}(u;m)\rt]=0,\label{Conservation}
\eea unitary condition, \bea
 R_{1,2}(u;m)\,R_{2,1}(-u;m)={\rm id}\otimes {\rm
 id},\label{Unitary}
\eea and crossing relation \bea
 R(u;m)^{kl}_{ij}=\varepsilon_{l}\,\varepsilon_{j}
   \frac{\sin(u)\sin((m-\eta\hat{\imath})_{21})}
   {\sin(u+\eta)\sin(m_{21})}R(-u-\eta;m-\eta\hat{\imath})
   ^{\bar{j}\,k}_{\bar{l}\,i},\label{Crossing}
\eea where
\bea \varepsilon_{1}=1,\,\varepsilon_{2}=-1,\quad {\rm
and}\,\, \bar{1}=2,\,\bar{2}=1.\label{Parity} \eea

Define the following functions: $\theta^{(1)}(u)=e^{-iu}$, $
\theta^{(2)}(u)=1$. Let us introduce two intertwiners which are
$2$-component  column vectors $\phi_{m,m-\eta\hat{\jmath}}(u)$
labelled by $\hat{1},\,\hat{2}$. The $k$-th element of
$\phi_{m,m-\eta\hat{\jmath}}(u)$ is given by \bea
\phi^{(k)}_{m,m-\eta\hat{\jmath}}(u)=\theta^{(k)}(u+2m_j).\label{Intvect}\eea
Explicitly, \bea \phi_{m,m-\eta\hat{1}}(u)=
\lt(\begin{array}{c}e^{-i(u+2m_1)}\\1\end{array}\rt),\qquad
\phi_{m,m-\eta\hat{2}}(u)=
\lt(\begin{array}{c}e^{-i(u+2m_2)}\\1\end{array}\rt).\eea
Obviously, the two intertwiner vectors
$\phi_{m,m-\eta\hat{\imath}}(u)$ are linearly {\it independent}
for a generic $m\in V$.

 Using the intertwiner vectors, one can derive the following face-vertex
correspondence relation \cite{Cao03}\bea &&\R_{1,2}(u_1-u_2)
\phi^1_{m,m-\eta\hat{\imath}}(u_1)
\phi^2_{m-\eta\hat{\imath},m-\eta(\hat{\imath}+\hat{\jmath})}(u_2)
\no\\&&~~~~~~= \sum_{k,l}R(u_1-u_2;m)^{kl}_{ij}
\phi^1_{m-\eta\hat{l},m-\eta(\hat{l}+\hat{k})}(u_1)
\phi^2_{m,m-\eta\hat{l}}(u_2). \label{Face-vertex} \eea  Then the
QYBE (\ref{QYB}) of the vertex-type R-matrix $\R(u)$ is equivalent
to the dynamical Yang-Baxter equation (\ref{MYBE}) of the SOS
R-matrix $R(u,m)$. For a generic $m$, we can introduce other types
of intertwiners $\bar{\phi},~\tilde{\phi}$ which  are both row
vectors and satisfy the following conditions, \bea
  &&\bar{\phi}_{m,m-\eta\hat{\mu}}(u)
     \,\phi_{m,m-\eta\hat{\nu}}(u)=\d_{\mu\nu},\quad
     \tilde{\phi}_{m+\eta\hat{\mu},m}(u)
     \,\phi_{m+\eta\hat{\nu},m}(u)=\d_{\mu\nu},\label{Int2}\eea
{}from which one  can derive the relations,
\begin{eqnarray}
&&\sum_{\mu=1}^2\phi_{m,m-\eta\hat{\mu}}(u)\,
 \bar{\phi}_{m,m-\eta\hat{\mu}}(u)={\rm id},\label{Int3}\\
&&\sum_{\mu=1}^2\phi_{m+\eta\hat{\mu},m}(u)\,
 \tilde{\phi}_{m+\eta\hat{\mu},m}(u)={\rm id}.\label{Int4}
\end{eqnarray}

One may verify that the K-matrices $K^{\pm}(u)$ given by
(\ref{K-matrix}) and (\ref{DK-matrix}) can be expressed in terms
of the intertwiners and {\it diagonal\/} matrices $\K(\l|u)$ and
$\tilde{\K}(\l|u)$ as follows \bea &&K^-(u)^s_t=
\sum_{i,j}\phi^{(s)}_{\l-\eta(\hat{\imath}-\hat{\jmath}),
~\l-\eta\hat{\imath}}(u)
\K(\l|u)^j_i\bar{\phi}^{(t)}_{\l,~\l-\eta\hat{\imath}}(-u),\label{K-F-1}\\
&&K^+(u)^s_t= \sum_{i,j}
\phi^{(s)}_{\l,~\l-\eta\hat{\jmath}}(-u)\tilde{\K}(\l|u)^j_i
\tilde{\phi}^{(t)}_{\l-\eta(\hat{\jmath}-\hat{\imath}),
~\l-\eta\hat{\jmath}}(u).\label{K-F-2}\eea Here the two {\it
diagonal\/} matrices $\K(\l|u)$ and $\tilde{\K}(\l|u)$ are given
by \bea
&&\K(\l|u)\equiv{\rm Diag}(k(\l|u)_1,\,k(\l|u)_2)={\rm
Diag}(\frac{\sin(\l_1+\xi-u)}{\sin(\l_1+\xi+u)},\,
\frac{\sin(\l_2+\xi-u)}{\sin(\l_2+\xi+u)}),\label{K-F-3}\\
&&\tilde{\K}(\l|u)\equiv{\rm
Diag}(\tilde{k}(\l|u)_1,\,\tilde{k}(\l|u)_2)\no\\
&&~~~~~~~~~={\rm
Diag}(\frac{\sin(\l_{12}\hspace{-0.1cm}-\hspace{-0.1cm}
\eta)\sin(\l_1\hspace{-0.1cm}+\hspace{-0.1cm}\bar{\xi}+\hspace{-0.1cm}u
\hspace{-0.1cm}+\hspace{-0.1cm}\eta)}
{\sin\l_{12}\sin(\l_1+\bar{\xi}-u-\eta)},\,
\frac{\sin(\l_{12}\hspace{-0.1cm}+\hspace{-0.1cm}
\eta)\sin(\l_2\hspace{-0.1cm}+\hspace{-0.1cm}\bar{\xi}\hspace{-0.1cm}
+\hspace{-0.1cm}u\hspace{-0.1cm}+\hspace{-0.1cm}\eta)}
{\sin\l_{12}\sin(\l_2+\bar{\xi}-u-\eta)}).\label{K-F-4} \eea
Although the vertex type K-matrices $K^{\pm}(u)$ given by
(\ref{K-matrix}) and (\ref{DK-matrix}) are generally non-diagonal,
after the face-vertex transformations (\ref{K-F-1}) and
(\ref{K-F-2}), the face type counterparts $\K(\l|u)$ and
$\tilde{\K}(\l|u)$  become {\it simultaneously\/} diagonal. This
fact enabled the authors to apply the generalized algebraic Bethe
ansatz method developed in \cite{Yan04} for SOS type integrable
models to diagonalize the transfer matrices $\t(u)$ (\ref{trans})
\cite{Yan04-1,Yan07}.

%%%%%%%%%%%%%%%%%%%%%%%%%%%%%%%%%%%%%%%%%%%%%%%%%%%%%%%%%%%%%%%%%%%%%%%%
\subsection{Two sets of eigenstates}

In order to construct the Bethe states of the open XXZ model
with non-diagonal boundary terms specified by the K-matrices
(\ref{K-matrix-2-1}) and (\ref{K-matrix-6}), we need to introduce
the new double-row
monodromy matrices $\T^{\pm}(m|u)$ \cite{Yan09}:
\bea
 \T^-(m|u)^{\nu}_{\mu}
     &=&\tilde{\phi}^{0}_{m-\eta(\hat{\mu}-\hat{\nu}),
        m-\eta\hat{\mu}}(u)~\mathbb{T}_0(u)\phi^{0}_{m,
        m-\eta\hat{\mu}}(-u),\label{Mon-F}\\
\T^+(m|u)^j_i
     &=&\prod_{k\neq j}\frac{\sin(m_{jk})}{\sin(m_{jk}-\eta)}
        \,\phi^{t_0}_{m-\eta(\hat{\jmath}-\hat{\imath}),m-\eta\hat{\jmath}}(u)
        \lt(\mathbb{T}^+(u)\rt)^{t_0}\bar{\phi}^{t_0}_{m,m-\eta\hat{\jmath}}(-u),
        \label{Mon-F-1}
\eea where $t_0$ denotes transposition in the $0$-th space (i.e.
auxiliary space) and $\mathbb{T}^+(u)$ is given by
\bea
  \lt(\mathbb{T}^+(u)\rt)^{t_0}&=&T^{t_0}(u)\lt(K^+(u)\rt)^{t_0}\hat{T}^{t_0}(u).
      \label{Mon-V-1}
\eea
These double-row monodromy matrices, in the face picture, can be
expressed in terms of the face type R-matrix $R(u;m)$
(\ref{R-matrix}) and  K-matrices  $\K(\l|u)$ (\ref{K-F-3}) and $\tilde{\K}(\l|u)$
(\ref{K-F-4}) (for the details see Appendix A).

So far  only two sets of Bethe states ( i.e. eigenstates) of the transfer matrix
for the models with non-diagonal boundary terms  have been found \cite{Yan07}.
These two sets of states are  \cite{Yan09}
\bea
&&|\{v^{(1)}_i\}\rangle^{(I)}=
     \T^+(\l+2\eta\hat{1}|v^{(1)}_1)^1_2\cdots
   \T^+(\l+2M\eta\hat{1}|v^{(1)}_M)^1_2|\O^{(I)}(\l)\rangle,
   \label{Bethe-state-1}\\
&&|\{v^{(2)}_i\}\rangle^{(II)} =
   \T^-(\l-2\eta\hat{2}|v^{(2)}_1)^2_1
   \cdots
   \T^-(\l\hspace{-0.04truecm}-\hspace{-0.04truecm}2M\eta\hat{2}|v^{(2)}_M)^2_1|\O^{(II)}(\l)\rangle,
   \label{Bethe-state-2}
\eea where the vector $\l$ is related to the boundary parameters
(\ref{boundary-vector}). The associated reference states
$|\O^{(I)}(\l)\rangle$ and $|\O^{(II)}(\l)\rangle$ are \bea
\hspace{-1.2truecm}|\O^{(I)}(\l)\rangle
   &=&\phi^1_{\l+N\eta\hat{1},\l+(N-1)\eta\hat{1}}(z_1)
      \phi^2_{\l+(N-1)\eta\hat{1},\l+(N-2)\eta\hat{1}}(z_{2})\cdots
      \phi^N_{\l+\eta\hat{1},\l}(z_N),\label{Vac-1}\\
\hspace{-1.2truecm} |\O^{(II)}(\l)\rangle&=&
\phi^1_{\l,\l-\eta\hat{2}}(z_1)
\phi^{2}_{\l-\eta\hat{2},\l-2\eta\hat{2}}(z_{2})\cdots
\phi^N_{\l-(N-1)\eta\hat{2},\l-N\eta\hat{2}}(z_N).\label{Vac-2}
\eea It is remarked that   $\phi^k={\rm id}\otimes {\rm
id}\cdots\otimes \stackrel{k-th}{\phi}\otimes {\rm id}\cdots$.

If the parameters $\{v^{(1)}_k\}$ satisfy the first set of  Bethe
ansatz equations given by
\bea &&\hspace{-0.1cm}\frac
{\sin(\l_2+\xi+v^{(1)}_{\a})\sin(\l_2+\bar\xi-v^{(1)}_{\a})
\sin(\l_1+\bar\xi+v^{(1)}_{\a})\sin(\l_1+\xi-v^{(1)}_{\a})}
{\sin(\l_2\hspace{-0.1cm}+\hspace{-0.1cm}
\bar\xi\hspace{-0.1cm}+\hspace{-0.1cm}v^{(1)}_{\a}
\hspace{-0.1cm}+\hspace{-0.1cm}\eta)
\sin(\l_2\hspace{-0.1cm}+\hspace{-0.1cm}\xi\hspace{-0.1cm}-\hspace{-0.1cm}v^{(1)}_{\a}
\hspace{-0.1cm}-\hspace{-0.1cm}\eta)
\sin(\l_1\hspace{-0.1cm}+\hspace{-0.1cm}\xi\hspace{-0.1cm}+\hspace{-0.1cm}
v^{(1)}_{\a}\hspace{-0.1cm}+\hspace{-0.1cm}\eta)
\sin(\l_1\hspace{-0.1cm}+\hspace{-0.1cm}\bar\xi\hspace{-0.1cm}-\hspace{-0.1cm}v^{(1)}_{\a}
\hspace{-0.1cm}-\hspace{-0.1cm}\eta)}\no\\
&&~~~~~~=\prod_{k\neq
\a}^M\frac{\sin(v^{(1)}_{\a}+v^{(1)}_k+2\eta)\sin(v^{(1)}_{\a}-v^{(1)}_k+\eta)}
{\sin(v^{(1)}_{\a}+v^{(1)}_k)\sin(v^{(1)}_{\a}-v^{(1)}_k-\eta)}\no\\
&&~~~~~~~~~~\times\prod_{k=1}^{2M}\frac{\sin(v^{(1)}_{\a}+z_k)\sin(v^{(1)}_{\a}-z_k)}
{\sin(v^{(1)}_{\a}+z_k+\eta)\sin(v^{(1)}_{\a}-z_k+\eta)},~~\a=1,\cdots,M,
\label{BA-D-1}\eea the Bethe state
$|v^{(I)}_1,\cdots,v^{(1)}_M\rangle^{(1)}$ becomes the eigenstate
of the transfer matrix with eigenvalue $\L^{(1)}(u)$  given by
\cite{Yan09}
\bea
&&\L^{(1)}(u)=\frac{\sin(\l_2+\bar\xi-u)\sin(\l_1+\bar\xi+u)\sin(\l_1+\xi-u)\sin(2u+2\eta)}
{\sin(\l_2+\bar\xi-u-\eta)\sin(\l_1+\bar\xi-u-\eta)\sin(\l_1+\xi+u)\sin(2u+\eta)}\no\\
&&~~~~~~~~~~~~~~~~~~\times\prod_{k=1}^M\frac{\sin(u+v^{(1)}_k)\sin(u-v^{(1)}_k-\eta)}
{\sin(u+v^{(1)}_k+\eta)\sin(u-v^{(1)}_k)}\no\\
&&~~~~~~+\frac{\sin(\l_2+\bar\xi+u+\eta)\sin(\l_1+\xi+u+\eta)\sin(\l_2+\xi-u-\eta)\sin
2u}
{\sin(\l_2+\bar\xi-u-\eta)\sin(\l_1+\xi+u)\sin(\l_2+\xi+u)\sin(2u+\eta)}\no\\
&&~~~~~~~~~~~~~~~~~~\times\prod_{k=1}^M\frac{\sin(u+v^{(1)}_k+2\eta)\sin(u-v^{(1)}_k+\eta)}
{\sin(u+v^{(1)}_k+\eta)\sin(u-v^{(1)}_k)}\no\\
&&~~~~~~~~~~~~~~~~~~\times\prod_{k=1}^{2M}\frac{\sin(u+z_k)\sin(u-z_k)}
{\sin(u+z_k+\eta)\sin(u-z_k+\eta)}.\label{Eigenfuction-D-1}
 \eea

\noindent If the parameters $\{v^{(2)}_k\}$ satisfy the second
Bethe Ansatz equations \bea
&&\hspace{-0.1cm}\frac
  {\sin(\l_1+\xi+v^{(2)}_{\a})\sin(\l_1+\bar\xi-v^{(2)}_{\a})
  \sin(\l_2+\bar\xi+v^{(2)}_{\a})\sin(\l_2+\xi-v^{(2)}_{\a})}
  {\sin(\l_1\hspace{-0.1cm}+\hspace{-0.1cm}
  \bar\xi\hspace{-0.1cm}+\hspace{-0.1cm}v^{(2)}_{\a}
  \hspace{-0.1cm}+\hspace{-0.1cm}\eta)
  \sin(\l_1\hspace{-0.1cm}+\hspace{-0.1cm}\xi\hspace{-0.1cm}-\hspace{-0.1cm}v^{(2)}_{\a}
  \hspace{-0.1cm}-\hspace{-0.1cm}\eta)
  \sin(\l_2\hspace{-0.1cm}+\hspace{-0.1cm}\xi\hspace{-0.1cm}+\hspace{-0.1cm}
  v^{(2)}_{\a}\hspace{-0.1cm}+\hspace{-0.1cm}\eta)
  \sin(\l_2\hspace{-0.1cm}+\hspace{-0.1cm}\bar\xi\hspace{-0.1cm}-\hspace{-0.1cm}v^{(2)}_{\a}
  \hspace{-0.1cm}-\hspace{-0.1cm}\eta)}\no\\
&&~~~~~~=\prod_{k\neq
  \a}^M\frac{\sin(v^{(2)}_{\a}+v^{(2)}_k+2\eta)\sin(v^{(2)}_{\a}-v^{(2)}_k+\eta)}
  {\sin(v^{(2)}_{\a}+v^{(2)}_k)\sin(v^{(2)}_{\a}-v^{(2)}_k-\eta)}\no\\
&&~~~~~~~~~~\times\prod_{k=1}^{2M}\frac{\sin(v^{(2)}_{\a}+z_k)\sin(v^{(2)}_{\a}-z_k)}
  {\sin(v^{(2)}_{\a}+z_k+\eta)\sin(v^{(2)}_{\a}-z_k+\eta)},~~\a=1,\cdots,M,
  \label{BA-D-2}
\eea the Bethe states $|v^{(2)}_1,\cdots,v^{(2)}_M\rangle^{(II)}$
yield the second set of the eigenstates of the transfer matrix
with the eigenvalues \cite{Yan07}, \bea
&&\L^{(2)}(u)=\frac{\sin(2u+2\eta)\sin(\l_1+\bar\xi-u)\sin(\l_2+\bar\xi+u)\sin(\l_2+\xi-u)}
{\sin(2u+\eta)\sin(\l_1+\bar\xi-u-\eta)\sin(\l_2+\bar\xi-u-\eta)\sin(\l_2+\xi+u)}\no\\
&&~~~~~~~~~~~~~~~~~~\times\prod_{k=1}^M\frac{\sin(u+v^{(2)}_k)\sin(u-v^{(2)}_k-\eta)}
{\sin(u+v^{(2)}_k+\eta)\sin(u-v^{(2)}_k)}\no\\
&&~~~~~~+\frac{\sin(2u)\sin(\l_1+\bar\xi+u+\eta)
\sin(\l_2+\xi+u+\eta)\sin(\l_1+\xi-u-\eta)}
{\sin(2u+\eta)\sin(\l_1+\bar\xi-u-\eta)\sin(\l_2+\xi+u)\sin(\l_1+\xi+u)}\no\\
&&~~~~~~~~~~~~~~~~~~\times\prod_{k=1}^M\frac{\sin(u+v^{(2)}_k+2\eta)\sin(u-v^{(2)}_k+\eta)}
{\sin(u+v^{(2)}_k+\eta)\sin(u-v^{(2)}_k)}\no\\
&&~~~~~~~~~~~~~~~~~~\times\prod_{k=1}^{2M}\frac{\sin(u+z_k)\sin(u-z_k)}
{\sin(u+z_k+\eta)\sin(u-z_k+\eta)}.\label{Eigenfuction-D-2}
 \eea

%%%%%%%%%%%%%%%%%%%%%%%%%%%%%%%%%%%%%%%%%%%%%%%%%%%%%%%%%%%%%%%
%                                                             %
%  3. Scalar   products                                       %
%                                                             %
%                                                             %
%                                                             %
%%%%%%%%%%%%%%%%%%%%%%%%%%%%%%%%%%%%%%%%%%%%%%%%%%%%%%%%%%%%%%%

\section{ Scalar products}
\label{T} \setcounter{equation}{0}

It was shown that in order to compute correlation functions of
the closed XXZ chain \cite{Kor93} and the open XXZ chain with
diagonal boundary terms \cite{Wan02,Kit07},
one suffices to calculate the scalar products
of an  on-shell Bethe state and a general state  (an off-shell Bethe state).
The aim of this paper is to give the explicit expressions of the following
scalar products  of the open XXZ chain with non-diagonal
boundary terms:
\bea
&&\hspace{-1.2truecm}S^{I,II}(\{u_{\a}\};\{v^{(2)}_i\})=
   {}^{(I)}\langle\{u_{\a}\}|\{v^{(2)}_i\}\rangle^{(II)},\quad
   S^{II,I}(\{u_{\a}\};\{v^{(1)}_i\})= {}^{(II)}\langle\{u_{\a}\}
   |\{v^{(1)}_i\}\rangle^{(I)},\label{Scalar-1}\\
&&\hspace{-1.2truecm}S^{I,I}(\{u_{\a}\};\{v^{(1)}_i\})=
   {}^{(I)}\langle\{u_{\a}\}|\{v^{(1)}_i\}\rangle^{(I)},\quad
   S^{II,II}(\{u_{\a}\};\{v^{(2)}_i\})=
   {}^{(II)}\langle\{u_{\a}\}|\{v^{(2)}_i\}\rangle^{(II)},\label{Scalar-2}
\eea where the dual states ${}^{(I)}\langle\{u_{\a}\}|$ and
${}^{(II)}\langle\{u_{\a}\}|$ are given by
\bea
&&{}^{(I)}\langle\{u_{\a}\}|=\langle\Omega^{(I)}(\l)|
   \T^-(\l-2(M-1)\eta\hat{1}|u_M)^2_1\ldots\T^-(\l|u_1)^2_1,\label{Dual-1}\\
&&{}^{(II)}\langle\{u_{\a}\}|=\langle\Omega^{(II)}(\l)|
   \T^+(\l+2(M-1)\eta\hat{2}|u_M)^1_2\ldots\T^+(\l|u_1)^1_2,\label{Dual-2}
\eea  and $\langle\Omega^{(I)}(\l)|$, $\langle\Omega^{(II)}(\l)|$ are
\bea
 && \langle\Omega^{(I)}(\l)|=\tilde{\phi}^1_{\l,\l-\eta\hat{1}}(z_1)\ldots
    \tilde{\phi}^N_{\l-(2M-1)\eta\hat{1},\l-2M\eta\hat{1}}(z_{N}),\\
 && \langle\Omega^{(II)}(\l)|=\tilde{\phi}^1_{\l+2M\eta\hat{2},\l+(2M-1)\eta\hat{2}}(z_1)\ldots
    \tilde{\phi}^N_{\l+\eta\hat{1},\l}(z_{N}).
\eea  Some remarks are in order. The parameters $\{u_{\a}\}$ in (\ref{Scalar-1})-(\ref{Scalar-2})
are free parameters, namely, they do not
need to satisfy the Bethe ansatz equations. However the parameters $\{v^{(1)}_{i}\}$ and
$\{v^{(2)}_{i}\}$ need to satisfy the Bethe ansatz equations (\ref{BA-D-1}) and (\ref{BA-D-2})
respectively.

The K-matrices $K^{\pm}(u)$ given by (\ref{K-matrix}) and
(\ref{DK-matrix}) are generally non-diagonal (in the vertex
picture), after the face-vertex transformations (\ref{K-F-1}) and
(\ref{K-F-2}), the face type counterparts $\K(\l|u)$ and
$\tilde{\K}(\l|u)$ given by (\ref{K-F-3}) and (\ref{K-F-4}) {\it
simultaneously\/} become diagonal. This fact suggests that it
would be much simpler if one performs all calculations in the face
picture.
%%%%%%%%%%%%%%%%%%%%%%%%%%%%%%%%%%%%%%%%%%%%%%%%%%%%%%%%%%%%%%%%%%%%%
\subsection{Face picture}

Let us introduce the face type one-row monodromy matrix (c.f
(\ref{Mon-V})) \bea
 T_{F}(l|u)&\equiv &T^{F}_{0,1\ldots N}(l|u)\no\\
 &=&R_{0,N}(u-z_N;l-\eta\sum_{i=1}^{N-1}h^{(i)})\ldots
    R_{0,2}(u-z_2;l-\eta h^{(1)})R_{0,1}(u-z_1;l),\no\\
 &=&\lt(\begin{array}{ll}T_F(l|u)^1_1&T_F(l|u)^1_2\\T_F(l|u)^2_1&
   T_F(l|u)^2_2\end{array}\rt)
    \label{Monodromy-face-1}
\eea where $l$ is a generic vector in $V$. The monodromy matrix
satisfies the face type quadratic exchange relation
\cite{Fel96,Hou03}. Applying $T_F(l|u)^i_j$ to an arbitrary vector
$|i_1,\ldots,i_N\rangle$ in the N-tensor product space $V^{\otimes
N}$ given by \bea
   |i_1,\ldots,i_N\rangle=\e^1_{i_1}\ldots
   \e^N_{i_N},\label{Vector-V}
\eea we have \bea
 T_F(l|u)^i_j|i_1,\ldots,i_N\rangle&\equiv&
    T_F(m;l|u)^i_j|i_1,\ldots,i_N\rangle\no\\
 &=&\sum_{\a_{N-1}\ldots\a_1}\sum_{i'_N\ldots i'_1}
 R(u-z_N;l-\eta\sum_{k=1}^{N-1}\hat{\imath}'_k)
   ^{i\,\,\,\,\,\,\,\,\,\,\,\,\,\,i'_N}_{\a_{N-1}\,i_N}\ldots\no\\
 &&\quad\quad\times R(u-z_2;l-\eta\hat{\imath}'_1)^{\a_2\,i'_2}_{\a_1\,\,i_2}
 R(u-z_1;l)^{\a_1\,i'_1}_{j\,\,\,\,i_1}
   \,\,|i'_1,\ldots,i'_N\rangle,\label{Monodromy-face-2}
\eea where $m=l-\eta\sum_{k=1}^N\hat{\imath}_k$. With the help of the crossing
relation (\ref{Crossing}), the face-vertex
correspondence relation (\ref{Face-vertex}) and the relations (\ref{Int2}),
following the
method developed in \cite{Yan04,Yan09}, we find that the scalar products (\ref{Scalar-1})-(\ref{Scalar-2})
can be expressed in terms of the face-type double-row monodromy operators as follows:
\bea
 &&S^{I,II}(\{u_{\a}\};\{v^{(2)}_i\})=\langle 1,\ldots,1|\T^-_F(\l-2(M-1)\eta\hat{1},\l|u_M)^2_1
     \ldots\T^-_F(\l,\l|u_1)^2_1 \no\\
     &&\qquad\qquad\times \T^-_F(\l+2\eta\hat{1},\l|v^{(2)}_1)^2_1 \ldots
     \T^-_F(\l+2M\eta\hat{1},\l|v^{(2)}_M)^2_1 |2,\ldots,2\rangle ,\label{Scalar-3}\\
 &&S^{II,I}(\{u_{\a}\};\{v^{(1)}_i\})=\langle 2,\ldots,2|\T^+_F(\l,\l+2(M-1)\eta\hat{2}|u_M)^1_2
     \ldots\T^+_F(\l,\l|u_1)^1_2 \no\\
     &&\qquad\qquad\times \T^+_F(\l,\l-2\eta\hat{2}|v^{(1)}_1)^1_2 \ldots
     \T^+_F(\l,\l-2M\eta\hat{2}|v^{(1)}_M)^1_2|1,\ldots,1\rangle,\\
 &&S^{I,I}(\{u_{\a}\};\{v^{(1)}_i\})=\langle 1,\ldots,1|\T^-_F(\l-2(M-1)\eta\hat{1},\l|u_M)^2_1
     \ldots\T^-_F(\l,\l|u_1)^2_1 \no\\
     &&\qquad\qquad\times \T^+_F(\l,\l+2\eta\hat{1}|v^{(1)}_1)^1_2 \ldots
     \T^+_F(\l,\l+2M\eta\hat{1}|v^{(1)}_M)^1_2|1,\ldots,1\rangle,\\
 &&S^{II,II}(\{u_{\a}\};\{v^{(2)}_i\})=\langle 2,\ldots,2|\T^+_F(\l,\l+2(M-1)\eta\hat{2}|u_M)^1_2
     \ldots\T^+_F(\l,\l|u_1)^1_2 \no\\
     &&\qquad\qquad\times \T^-_F(\l-2\eta\hat{2},\l|v^{(2)}_1)^2_1 \ldots
     \T^-_F(\l-2M\eta\hat{2},\l|v^{(2)}_M)^2_1 |2,\ldots,2\rangle.\label{Scalar-4}
\eea  The above double-row monodromy  matrix operators $\T^-_F(m,\l|u)^2_1$ and  $\T^+_F(\l,m|u)^1_2$ are given
in terms of the one-row monodromy matrix operator $T_F(m;l|u)^i_j$
\cite{Yan09}
\bea
 &&\T^-_F(m,\l|u)^2_1=
 \frac{\sin(m_{21})}{\sin(\l_{21})}\prod_{k=1}^N
      \frac{\sin(u+z_k)}{\sin(u+z_k+\eta)}\no\\
 &&\,\quad\times\lt\{
      \frac{\sin(\l_1+\xi-u)}{\sin(\l_1+\xi+u)}
      T_F(m,\l|u)^2_1
      T_F(m+\eta\hat{2},\l+\eta\hat{2}|-u-\eta)^2_2\rt.\no\\
 &&\,\qquad-\lt.
      \frac{\sin(\l_2+\xi-u)}{\sin(\l_2+\xi+u)}
      T_F(m+2\eta\hat{2},\l|u)^2_2
      T_F(m+\eta\hat{1},\l+\eta\hat{1}|-u-\eta)^2_1\rt\},\label{Expression-3}\\
 &&\T^+_F(\l,m|u)^1_2=\prod_{k=1}^N
      \frac{\sin(u+z_k)}{\sin(u+z_k+\eta)}\no\\
 &&\,\quad\times\lt\{
      \frac{\sin(\l_{12}\hspace{-0.08truecm}-\hspace{-0.08truecm}\eta)
      \sin(\l_1\hspace{-0.08truecm}+\hspace{-0.08truecm}\bar{\xi}
      \hspace{-0.08truecm}+\hspace{-0.08truecm}u\hspace{-0.08truecm}+\hspace{-0.08truecm}\eta)}
      {\sin(m_{12}\hspace{-0.08truecm}-\hspace{-0.08truecm}\eta)
      \sin(\l_1\hspace{-0.08truecm}+\hspace{-0.08truecm}\bar{\xi}\hspace{-0.08truecm}-
      \hspace{-0.08truecm}u\hspace{-0.08truecm}-\hspace{-0.08truecm}\eta)}
      T_F(\l\hspace{-0.08truecm}+\hspace{-0.08truecm}2\eta\hat{2},m
      \hspace{-0.08truecm}+\hspace{-0.08truecm}2\eta\hat{2}|u)^1_2
      T_F(\l\hspace{-0.08truecm}+\hspace{-0.08truecm}\eta\hat{2},m
      \hspace{-0.08truecm}+\hspace{-0.08truecm}\eta\hat{2}|
      \hspace{-0.08truecm}-\hspace{-0.08truecm}u
      \hspace{-0.08truecm}-\hspace{-0.08truecm}\eta)^2_2\rt.\no\\
 &&\,\qquad-\lt.
      \frac{\sin(\l_{21}\hspace{-0.08truecm}-\hspace{-0.08truecm}\eta)
      \sin(\l_2\hspace{-0.08truecm}+\hspace{-0.08truecm}\bar{\xi}
      \hspace{-0.08truecm}+\hspace{-0.08truecm}u
      \hspace{-0.08truecm}+\hspace{-0.08truecm}\eta)}
      {\sin(m_{21}\hspace{-0.08truecm}+\hspace{-0.08truecm}\eta)
      \sin(\l_2\hspace{-0.08truecm}+\hspace{-0.08truecm}\bar{\xi}
      \hspace{-0.08truecm}-\hspace{-0.08truecm}u
      \hspace{-0.08truecm}-\hspace{-0.08truecm}\eta)}
      T_F(\l,m\hspace{-0.08truecm}+\hspace{-0.08truecm}2\eta\hat{2}|u)^2_2
      T_F(\l\hspace{-0.08truecm}+\hspace{-0.08truecm}\eta\hat{2},m
      \hspace{-0.08truecm}+\hspace{-0.08truecm}\eta\hat{2}|
      \hspace{-0.08truecm}-\hspace{-0.08truecm}u\hspace{-0.08truecm}-\hspace{-0.08truecm}
      \eta)^1_2\rt\}.\no\\
 &&\label{Expression-4}
\eea

In the next section we shall construct the Drinfeld twist (or
factorizing F-matrix) in the face picture for the open XXZ chain
with non-diagonal boundary terms. In the resulting  F-basis, the two sets of
pseudo-particle creation/annihilation operators $\T^{\pm}_F$ given by
(\ref{Expression-3}) and (\ref{Expression-4}) take  completely
symmetric and polarization free forms simultaneously. These polarization free forms
allow us to construct the explicit expressions of the scalar products (\ref{Scalar-3})-
(\ref{Scalar-4}).

%%%%%%%%%%%%%%%%%%%%%%%%%%%%%%%%%%%%%%%%%%%%%%%%%%%%%%%%%%%%%%%
%                                                             %
%  4. F-basis                                                 %
%                                                             %
%                                                             %
%                                                             %
%%%%%%%%%%%%%%%%%%%%%%%%%%%%%%%%%%%%%%%%%%%%%%%%%%%%%%%%%%%%%%%

\section{ F-basis}
\label{F-basis} \setcounter{equation}{0}

In this section, we construct the Drinfeld twist \cite{Dri83}
(factorizing F-matrix) on the $N$-fold tensor product space
$V^{\otimes N}$ (i.e. the quantum space of the open XXZ chain) and
the associated representations of the pseudo-particle creation/annihilation
operators in this basis.

\subsection{Factorizing Drinfeld twist $F$}
Let $ \mathcal{S}_N$ be the permutation group over indices
$1,\ldots,N$ and $\{\s_i|i=1,\ldots,N-1\}$ be the set of
elementary permutations in $\mathcal{S}_N$. For each elementary
permutation $\s_i$, we introduce the associated operator
$R^{\s_i}_{1\ldots N}$ on the quantum space \bea
  R^{\s_i}_{1\ldots N}(l)\equiv R^{\s_i}(l)=R_{i,i+1}
    (z_i-z_{i+1}|l-\eta\sum_{k=1}^{i-1}h^{(k)}),\label{Fundamental-R-operator}
\eea where $l$ is a generic vector in $V$. For any $\s,\,\s'\in
\mathcal{S}_N$, operator $R^{\s\s'}_{1\ldots N}$ associated with
$\s\s'$ satisfies the following composition law \cite{Yan09}(and references therein):
\bea
  R_{1\ldots N}^{\s\s'}(l)=R^{\s'}_{\s(1\ldots
  N)}(l)\,R^{\s}_{1\ldots N}(l).\label{Rule}
\eea Let $\s$ be decomposed in a minimal way in terms of
elementary permutations,
\bea
  \s=\s_{\b_1}\ldots\s_{\b_p}, \label{decomposition}
\eea where $\b_i=1,\ldots, N-1$ and the positive integer $p$ is
the length of $\s$. The composition law (\ref{Rule}) enables one
to obtain  operator $R^{\s}_{1\ldots N}$ associated with each
$\s\in\mathcal{S}_N $. The dynamical quantum Yang-Baxter equation
(\ref{MYBE}), weight conservation condition (\ref{Conservation})
and unitary condition (\ref{Unitary}) guarantee the uniqueness of
$R^{\s}_{1\ldots N}$. Moreover, one may check that
$R^{\s}_{1\ldots N}$ satisfies the following exchange relation
with the face type one-row monodromy matrix
(\ref{Monodromy-face-1}) \bea
  R^{\s}_{1\ldots N}(l)T^F_{0,1\ldots N}(l|u)=T^F_{0,\s(1\ldots N)}(l|u)
    R^{\s}_{1\ldots N}(l-\eta h^{(0)}),\quad\quad \forall\s\in
    \mathcal{S}_N.\label{Exchang-Face-1}
\eea

Now, we construct the face-type Drinfeld twist $F_{1\ldots
N}(l)\equiv F_{1\ldots N}(l;z_1,\ldots,z_N)$ \footnote{In this
paper, we adopt the convention: $F_{\s(1\ldots N)}(l)\equiv
F_{\s(1\ldots N)}(l;z_{\s(1)},\ldots,z_{\s(N)})$.} on the $N$-fold
tensor product space $V^{\otimes N}$, which  satisfies the
following three properties:
\bea
 &&{\rm I.\,\,\,\,lower-triangularity;}\\
 &&{\rm II.\,\,\, non-degeneracy;}\\
 &&{\rm III.\,factorizing \, property}:\,\,
 R^{\s}_{1\ldots N}(l)\hspace{-0.08truecm}=\hspace{-0.08truecm}
    F^{-1}_{\s(1\ldots N)}(l)F_{1\ldots N}(l), \,\,
 \forall\s\in  \mathcal{S}_N.\label{Factorizing}
\eea Substituting (\ref{Factorizing}) into the exchange relation
(\ref{Exchang-Face-1}) yields the following relation
\bea
 F^{-1}_{\s(1\ldots N)}(l)F_{1\ldots N}(l)T^F_{0,1\ldots N}(l|u)=
   T^F_{0,\s(1\ldots N)}(l|u)F^{-1}_{\s(1\ldots N)}(l-\eta h^{(0)})
   F_{1\ldots N}(l-\eta h^{(0)}).
\eea Equivalently,
\bea
 F_{1\ldots N}(l)T^F_{0,1\ldots N}(l|u)F^{-1}_{1\ldots N}(l-\eta h^{(0)})
   =F_{\s(1\ldots N)}(l)T^F_{0,\s(1\ldots N)}(l|u)
   F^{-1}_{\s(1\ldots N)}(l-\eta h^{(0)}).\label{Invariant}
\eea Let us introduce the twisted monodromy matrix
$\tilde{T}^F_{0,1\ldots N}(l|u)$ by \bea
 \tilde{T}^F_{0,1\ldots N}(l|u)&=&
  F_{1\ldots N}(l)T^F_{0,1\ldots N}(l|u)F^{-1}_{1\ldots N}(l-\eta
  h^{(0)})\no\\
  &=&\lt(\begin{array}{ll}\tilde{T}_F(l|u)^1_1&\tilde{T}_F(l|u)^1_2
  \\\tilde{T}_F(l|u)^2_1&
   \tilde{T}_F(l|u)^2_2\end{array}\rt).\label{Twisted-Mon-F}
\eea Then (\ref{Invariant}) implies that the twisted monodromy
matrix is symmetric under $\mathcal{S}_N$, namely, \bea
 \tilde{T}^F_{0,1\ldots N}(l|u)=\tilde{T}^F_{0,\s(1\ldots
 N)}(l|u), \quad \forall \s\in \mathcal{S}_N.
\eea

Define the F-matrix:
\bea
  F_{1\ldots N}(l)=\sum_{\s\in
     \mathcal{S}_N}\sum^2_{\{\a_j\}=1}\hspace{-0.22truecm}{}^*\,\,\,\,
     \prod_{j=1}^NP^{\s(j)}_{\a_{\s(j)}}
     \,R^{\s}_{1\ldots N}(l),\label{F-matrix}
\eea where $P^i_{\a}$ is the embedding of the project operator
$P_{\a}$ in the $i^{{\rm th}}$ space with matric elements
$(P_{\a})_{kl}=\d_{kl}\d_{k\a}$. The sum $\sum^*$ in
(\ref{F-matrix}) is over all non-decreasing sequences of the
labels $\a_{\s(i)}$:
\bea
  && \a_{\s(i+1)}\geq \a_{\s(i)}\quad {\rm if}\quad \s(i+1)>\s(i),\no\\
  && \a_{\s(i+1)}> \a_{\s(i)}\quad {\rm if}\quad
  \s(i+1)<\s(i).\label{Condition}
\eea From (\ref{Condition}), $F_{1\ldots N}(l)$ obviously is a
lower-triangular matrix. Moreover, the F-matrix is non-degenerate
because  all its diagonal elements are non-zero. It was shown \cite{Yan09}
that the F-matrix also satisfies the
factorizing property (\ref{Factorizing}). Hence, the F-matrix
$F_{1\ldots N}(l)$ given by (\ref{F-matrix}) is the desirable
Drinfeld twist.

\subsection{Completely symmetric  representations}
Direct calculation shows \cite{Yan09}  that the
twisted operators $\tilde{T}_F(l|u)^j_i$ defined by
(\ref{Twisted-Mon-F}) indeed simultaneously  have the
following polarization free forms
\bea
 &&\tilde{T}_F(l|u)^2_2=\frac{\sin(l_{21}-\eta)}{\sin\lt(l_{21}-\eta+
     \eta\langle H,\e_1\rangle\rt)}\otimes_{i}
     \lt(\begin{array}{ll}\frac{\sin(u-z_i)}{\sin(u-z_i+\eta)}&\\
     &1\end{array}\rt)_{(i)},\\[6pt]
 &&\tilde{T}_F(l|u)^2_1=\sum_{i=1}^N\frac{\sin\eta
     \sin(u\hspace{-0.08truecm}-\hspace{-0.08truecm}z_i
     \hspace{-0.08truecm}+\hspace{-0.08truecm}l_{12})}
     {\sin(u\hspace{-0.08truecm}-\hspace{-0.08truecm}z_i
     \hspace{-0.08truecm}+\hspace{-0.08truecm}\eta)\sin l_{12}} E_{12}^i\otimes_{j\neq i}
     \lt(\begin{array}{ll}\frac{\sin(u-z_j)\sin(z_i-z_j+\eta)}{\sin(u-z_j+\eta)\sin(z_i-z_j)}&\\
     &1\end{array}\rt)_{(j)},\\[6pt]
 &&\tilde{T}_F(l|u)^1_2=\frac{\sin(l_{21}\hspace{-0.08truecm}-\hspace{-0.08truecm}\eta)}
     {\sin(l_{21}\hspace{-0.08truecm}+\hspace{-0.08truecm}
     \eta\langle H,\e_1\hspace{-0.08truecm}-\hspace{-0.08truecm}\e_2\rangle)}
     \hspace{-0.08truecm}
     \sum_{i=1}^N\frac{\sin\eta\sin(u\hspace{-0.08truecm}-\hspace{-0.08truecm}z_i
     \hspace{-0.08truecm}+\hspace{-0.08truecm}l_{21}\hspace{-0.08truecm}
     +\hspace{-0.08truecm}\eta\hspace{-0.08truecm}
     +\hspace{-0.08truecm}\eta\langle H,\e_1
     \hspace{-0.08truecm}-\hspace{-0.08truecm}\e_2\rangle)}
     {\sin(u\hspace{-0.08truecm}-\hspace{-0.08truecm}z_i\hspace{-0.08truecm}+\hspace{-0.08truecm}\eta)
     \sin(l_{21}\hspace{-0.08truecm}+\hspace{-0.08truecm}\eta
     \hspace{-0.08truecm}+\hspace{-0.08truecm}\eta\langle
     H,\e_1\hspace{-0.08truecm}-\hspace{-0.08truecm}\e_2\rangle)}\no\\
 &&\quad\quad\quad\quad\quad\quad
     \times E_{21}^i\otimes_{j\neq i} \lt( \begin{array}{ll}
     \frac{\sin(u-z_j)}{\sin(u-z_j+\eta)}&\\
     &\frac{\sin(z_j-z_i+\eta)}{\sin(z_j-z_i)}\end{array}
     \rt)_{(j)},
\eea where $H=\sum_{k=1}^N h^{(k)}$. Applying  the above operators
to the arbitrary  state $|i_1,\ldots,i_N\rangle$ given by
(\ref{Vector-V}) leads to
\bea
 &&\tilde{T}_F(m,l|u)^2_2=\frac{\sin(l_{21}-\eta)}{\sin\lt(l_{2}-m_1-\eta\rt)}
     \otimes_{i}
     \lt(\begin{array}{ll}\frac{\sin(u-z_i)}{\sin(u-z_i+\eta)}&\\
     &1\end{array}\rt)_{(i)},\\[6pt]
 &&\tilde{T}_F(m,l|u)^2_1=\sum_{i=1}^N
     \frac{\sin\eta
     \sin(u-z_i+l_{12})}{\sin(u-z_i+\eta)\sin l_{12}}\no\\
 &&\quad\quad\quad\quad\quad\quad
     \times    E_{12}^i \otimes_{j\neq i}
     \lt(\begin{array}{ll}\frac{\sin(u-z_j)\sin(z_i-z_j+\eta)}{\sin(u-z_j+\eta)\sin(z_i-z_j)}&\\
     &1\end{array}\rt)_{(j)},\\[6pt]
 &&\tilde{T}_F(m,l|u)^1_2=\frac{\sin(l_{21}-\eta)}
     {\sin(m_{21}-2\eta)}
     \sum_{i=1}^N\frac{\sin\eta\sin(u-z_i+m_{21}-\eta)}
     {\sin(u-z_i+\eta)\sin(m_{21}-\eta)}\no\\
 &&\quad\quad\quad\quad\quad\quad
     \times E_{21}^i\otimes_{j\neq i} \lt( \begin{array}{ll}
     \frac{\sin(u-z_j)}{\sin(u-z_j+\eta)}&\\
     &\frac{\sin(z_j-z_i+\eta)}{\sin(z_j-z_i)}\end{array}
     \rt)_{(j)}.
\eea It then follows that the two pseudo-particle creation
operators (\ref{Expression-3}) and (\ref{Expression-4}) in the
F-basis simultaneously have the following completely symmetric
polarization free forms:
\bea
 &&\tilde{\T}^-_F(m,\l|u)^2_1=\frac{\sin m_{12}}{\sin(m_1-\l_2)}
   \prod_{k=1}^N\frac{\sin(u+z_k)}{\sin(u+z_k+\eta)}\no\\
  &&\quad\quad\times \sum_{i=1}^N\frac{\sin(\l_1+\xi-z_i)\sin(\l_2+\xi+z_i)\sin2u \sin\eta}
   {\sin(\l_1+\xi+u)\sin(\l_2+\xi+u)\sin(u-z_i+\eta)\sin(u+z_i)}\no\\[6pt]
  &&\quad\quad\quad\quad\quad\quad \times
   E_{12}^i\otimes_{j\neq i}\lt(\begin{array}{ll}
   \frac{\sin(u-z_j)\sin(u+z_j+\eta)\sin(z_i-z_j+\eta)}
   {\sin(u-z_j+\eta)\sin(u+z_j)\sin(z_i-z_j)}&\\
   &1\end{array}\rt)_{(j)},\label{Creation-operator-1}\\[6pt]
 &&\tilde{\T}^+_F(\l,m|u)^1_2=\frac{\sin (m_{21}+\eta)}{\sin(m_2-\l_1)}
   \prod_{k=1}^N\frac{\sin(u+z_k)}{\sin(u+z_k+\eta)}\no\\
  &&\quad\quad\times \sum_{i=1}^N
   \hspace{-0.08truecm}
   \frac{\sin(\l_2\hspace{-0.08truecm}+\hspace{-0.08truecm}\bar{\xi}
   \hspace{-0.08truecm}-\hspace{-0.08truecm}z_i)
   \sin(\l_1\hspace{-0.08truecm}+\hspace{-0.08truecm}\bar{\xi}
   \hspace{-0.08truecm}+\hspace{-0.08truecm}z_i)
   \sin(2u\hspace{-0.08truecm}+\hspace{-0.08truecm}2\eta) \sin\eta}
   {\sin(\l_1\hspace{-0.08truecm}+\hspace{-0.08truecm}\bar{\xi}
   \hspace{-0.08truecm}-\hspace{-0.08truecm}u
   \hspace{-0.08truecm}-\hspace{-0.08truecm}\eta)
   \sin(\l_2\hspace{-0.08truecm}+\hspace{-0.08truecm}\bar{\xi}
   \hspace{-0.08truecm}-\hspace{-0.08truecm}u
   \hspace{-0.08truecm}-\hspace{-0.08truecm}\eta)
   \sin(u\hspace{-0.08truecm}+\hspace{-0.08truecm}z_i)
   \sin(u\hspace{-0.08truecm}-\hspace{-0.08truecm}z_i
   \hspace{-0.08truecm}+\hspace{-0.08truecm}\eta)}\no\\[6pt]
  &&\quad\quad\quad\quad\quad\quad \times
   E_{21}^i\otimes_{j\neq i}\lt(\begin{array}{ll}
   \frac{\sin(u-z_j)\sin(u+z_j+\eta)}
   {\sin(u-z_j+\eta)\sin(u+z_j)}&\\
   &\frac{\sin(z_j-z_i+\eta)}{\sin(z_j-z_i)}\end{array}\rt)_{(j)}.\label{Creation-operator-2}
\eea

%%%%%%%%%%%%%%%%%%%%%%%%%%%%%%%%%%%%%%%%%%%%%%%%%%%%%%%%%%%%%%%
%                                                             %
%  5. Determinant representations of the scalar products      %
%                                                             %
%                                                             %
%                                                             %
%%%%%%%%%%%%%%%%%%%%%%%%%%%%%%%%%%%%%%%%%%%%%%%%%%%%%%%%%%%%%%%

\section{Determinant representations of the scalar products}
\label{F} \setcounter{equation}{0}

Due to the fact that the states $|1,\ldots,1\rangle$, $|2,\ldots,2\rangle$ and their dual states
$\langle 1,\ldots,1|$, $\langle 2,\ldots,2|$ are
invariant under the action of the F-matrix $F_{1\ldots N}(l)$
(\ref{F-matrix}), the calculation of the scalar products (\ref{Scalar-3})-(\ref{Scalar-4})
can be performed in the F-basis. Namely,
\bea
 &&S^{I,II}(\{u_{\a}\};\{v^{(2)}_i\})=\langle 1,\ldots,1|\tilde{\T}^-_F(\l-2(M-1)\eta\hat{1},\l|u_M)^2_1
     \ldots\tilde{\T}^-_F(\l,\l|u_1)^2_1 \no\\
     &&\qquad\qquad\times \tilde{\T}^-_F(\l+2\eta\hat{1},\l|v^{(2)}_1)^2_1 \ldots
     \tilde{\T}^-_F(\l+2M\eta\hat{1},\l|v^{(2)}_M)^2_1 |2,\ldots,2\rangle ,\label{Scalar-5}\\
 &&S^{II,I}(\{u_{\a}\};\{v^{(1)}_i\})=\langle 2,\ldots,2|\tilde{\T}^+_F(\l,\l+2(M-1)\eta\hat{2}|u_M)^1_2
     \ldots\tilde{\T}^+_F(\l,\l|u_1)^1_2 \no\\
     &&\qquad\qquad\times \tilde{\T}^+_F(\l,\l-2\eta\hat{2}|v^{(1)}_1)^1_2 \ldots
     \tilde{\T}^+_F(\l,\l-2M\eta\hat{2}|v^{(1)}_M)^1_2|1,\ldots,1\rangle,\\
 &&S^{I,I}(\{u_{\a}\};\{v^{(1)}_i\})=\langle 1,\ldots,1|\tilde{\T}^-_F(\l-2(M-1)\eta\hat{1},\l|u_M)^2_1
     \ldots\tilde{\T}^-_F(\l,\l|u_1)^2_1 \no\\
     &&\qquad\qquad\times \tilde{\T}^+_F(\l,\l+2\eta\hat{1}|v^{(1)}_1)^1_2 \ldots
     \tilde{\T}^+_F(\l,\l+2M\eta\hat{1}|v^{(1)}_M)^1_2|1,\ldots,1\rangle,\label{Scalar-6}\\
 &&S^{II,II}(\{u_{\a}\};\{v^{(2)}_i\})=\langle 2,\ldots,2|\tilde{\T}^+_F(\l,\l+2(M-1)\eta\hat{2}|u_M)^1_2
     \ldots\tilde{\T}^+_F(\l,\l|u_1)^1_2 \no\\
     &&\qquad\qquad\times \tilde{\T}^-_F(\l-2\eta\hat{2},\l|v^{(2)}_1)^2_1 \ldots
     \tilde{\T}^-_F(\l-2M\eta\hat{2},\l|v^{(2)}_M)^2_1 |2,\ldots,2\rangle.\label{Scalar-7}
\eea In the above equations, we have used the identity: $\hat{1}=-\hat{2}$.
Thanks to the polarization free representations
(\ref{Creation-operator-1}) and (\ref{Creation-operator-2}) of the
pseudo-particle creation/annihilation  operators,  we  can obtain the determinant representations
of the scalar products.

%%%%%%%%%%%%%%%%%%%%%%%%%%%%%%%%%%%%%%%%%%%%%%%%%%%%%%%%%%%%%%%%%%%%%%%%%%%
\subsection{The scalar products $S^{I,II}$ and $S^{II,I}$}
It was shown \cite{Yan10} that the scalar product $S^{I,II}(\{u_{\a}\};\{v^{(2)}_i\})$ (resp.
$S^{II,I}(\{u_{\a}\};\{v^{(1)}_i\})$) can be expressed in terms of some determinant no matter the
parameters $\{v^{(2)}_i\}$ (resp.$\{v^{(1)}_i\}$ ) satisfy the associated Bethe ansatz equations or not. In this subsection
we do not require these parameters being the roots of the Bethe ansatz equations. Let us introduce two functions
\bea
 {\cal Z}^{(I)}_N(\{\bar{u}_{J}\})\hspace{-0.36truecm}&\equiv&\hspace{-0.36truecm} S^{I,II}(\{u_{\a}\};\{v_i\})\no\\
     &=&\hspace{-0.36truecm} \langle 1,\ldots,1|\tilde{\T}^-_F
     (\l\hspace{-0.1truecm}-\hspace{-0.1truecm}2(M\hspace{-0.1truecm}-\hspace{-0.1truecm}1)
     \eta\hat{1},\l|\bar{u}_{N})^2_1 \ldots
     \tilde{\T}^-_F(\l\hspace{-0.1truecm}+\hspace{-0.1truecm}2M\eta\hat{1},\l|\bar{u}_1)^2_1 |2,\ldots,2\rangle,\\
 {\cal Z}^{(II)}_N(\{\bar{u}_{J}\})\hspace{-0.36truecm}&\equiv& \hspace{-0.36truecm}
     S^{II,I}(\{u_{\a}\};\{v_i\})\no\\
     &=&\hspace{-0.36truecm}\langle 2,\ldots,2|\tilde{\T}^+_F(\l,\l\hspace{-0.1truecm}+\hspace{-0.1truecm}
     2(M\hspace{-0.1truecm}-\hspace{-0.1truecm}1)\eta\hat{2}|\bar{u}_{N})^1_2\ldots
     \tilde{\T}^+_F(\l,\l\hspace{-0.1truecm}-\hspace{-0.1truecm}2M\eta\hat{2}|\bar{u}_1)^1_2|1,\ldots,1\rangle,
\eea where $N$ free parameters $\{\bar{u}_J|J=1,\ldots N\}$ are given by
\bea
\bar{u}_i=u_i\,\, {\rm for}\,\, i=1,\ldots M,\qquad {\rm and}\qquad\bar{u}_{M+i}=v_i \,\, {\rm for}\,\, i=1,\ldots M.
\eea  Note that these
functions ${\cal Z}^{(I)}_N(\{\bar{u}_{J}\})$ and ${\cal Z}^{(II)}_N(\{\bar{u}_{J}\})$ correspond to the partition functions
of the six-vertex model with domain wall
boundary conditions and one reflecting end \cite{Tsu98} specified by the non-diagonal K-matrices
(\ref{K-matrix}) and (\ref{DK-matrix})
respectively \cite{Yan10}.

The polarization free representations
(\ref{Creation-operator-1}) and (\ref{Creation-operator-2}) of the
pseudo-particle creation/annihilation operators allowed ones \cite{Yan10} to express the above functions in terms of the
determinants representations of some $N\times N$ matrices as follows:
\bea
{\cal Z}^{(I)}_N(\{\bar{u}_{J}\})&=&\prod_{k=1}^M\frac{\sin(\l_{12}+2k\eta)\sin(\l_{12}-2k\eta+\eta)}
  {\sin(\l_{12}+k\eta)\sin(\l_{12}-k\eta+\eta)}
  \prod_{l=1}^N\prod_{i=1}^N\frac{\sin(\bar{u}_i+z_l)}{\sin(\bar{u}_i+z_l+\eta)}\no\\
  &&\quad \times \frac{\prod_{\a=1}^N\prod_{i=1}^N\sin(\bar{u}_{\a}-z_i)\sin(\bar{u}_{\a}+z_i+\eta)
   {\rm det}{\cal N}^{(I)}(\{\bar{u}_{\a}\};\{z_i\})}
  {\prod_{\a>\b}\sin(\bar{u}_{\a}\hspace{-0.1truecm}-\hspace{-0.1truecm}
  \bar{u}_{\b})\sin(\bar{u}_{\a}\hspace{-0.1truecm}+\hspace{-0.1truecm}\bar{u}_{\b}
  \hspace{-0.1truecm}+\hspace{-0.1truecm}\eta)\prod_{k<l}
  \sin(z_k\hspace{-0.1truecm}-\hspace{-0.1truecm}z_l)\sin(z_k\hspace{-0.1truecm}+\hspace{-0.1truecm}z_l)},
  \label{partition-1}\\
{\cal Z}^{(II)}_N(\{\bar{u}_{J}\})&=&\prod_{k=1}^M\frac{\sin(\l_{21}+\eta-2k\eta)\sin(\l_{21}-\eta+2k\eta)}
  {\sin(\l_{21}-k\eta)\sin(\l_{21}+k\eta-\eta)}
  \prod_{l=1}^N\prod_{i=1}^N\frac{\sin(\bar{u}_i+z_l)}{\sin(\bar{u}_i+z_l+\eta)}\no\\
  &&\quad \times \frac{\prod_{\a=1}^N\prod_{i=1}^N\sin(\bar{u}_{\a}+z_i)\sin(\bar{u}_{\a}-z_i+\eta)
   {\rm det}{\cal N}^{(II)}(\{\bar{u}_{\a}\};\{z_i\})}
  {\prod_{\a>\b}\sin(\bar{u}_{\a}\hspace{-0.1truecm}-\hspace{-0.1truecm}
  \bar{u}_{\b})\sin(\bar{u}_{\a}\hspace{-0.1truecm}+\hspace{-0.1truecm}\bar{u}_{\b}
  \hspace{-0.1truecm}+\hspace{-0.1truecm}\eta)\prod_{k<l}
  \sin(z_l\hspace{-0.1truecm}-\hspace{-0.1truecm}z_k)\sin(z_l\hspace{-0.1truecm}+\hspace{-0.1truecm}z_k)},
  \label{partition-2}
\eea where the $N\times N$ matrices  ${\cal N}^{(I)}(\{\bar{u}_{\a}\};\{z_i\})$ and
${\cal N}^{(II)}(\{\bar{u}_{\a}\};\{z_i\})$ are given by
\bea
{\cal N}^{(I)}(\{\bar{u}_{\a}\};\{z_i\})_{\a,j}&=&
  \frac{\sin\eta\sin(\l_1+\xi-z_j)}
  {\sin(\bar{u}_{\a}-z_j)\sin(\bar{u}_{\a}+z_j+\eta)
  \sin(\l_1+\xi+\bar{u}_{\a})}\no\\
  &&\times \frac{\sin(\l_2+\xi+z_j)\sin(2\bar{u}_{\a})}
  {\sin(\l_2+\xi+\bar{u}_{\a})
  \sin(\bar{u}_{\a}-z_j+\eta)
  \sin(\bar{u}_{\a}+z_j)},\\
{\cal N}^{(II)}(\{\bar{u}_{\a}\};\{z_i\})_{\a,j}&=&
  \frac{\sin\eta\sin(\l_2+\bar{\xi}-z_j)}
  {\sin(\bar{u}_{\a}-z_j)\sin(\bar{u}_{\a}+z_j+\eta)
  \sin(\l_2+\bar{\xi}-\bar{u}_{\a}-\eta)}\no\\
  &&\times \frac{\sin(\l_1+\bar{\xi}+z_j)\sin(2\bar{u}_{\a}+2\eta)}
  {\sin(\l_1+\bar{\xi}-\bar{u}_{\a}-\eta)
  \sin(\bar{u}_{\a}-z_j+\eta)
  \sin(\bar{u}_{\a}+z_j)}.
\eea  The above determinant representations are crucial to construct
the determinant representations of the remaining scalar products $S^{I,I}$ and $S^{II,II}$ in the
next subsection.

\subsection{The scalar products $S^{I,I}$ and $S^{II,II}$}
Let us introduce two sets of  functions $\{H^{(I)}_j(u;\{z_i\},\{v_i\})|j=1,\ldots,M\}$ and
$\{H^{(II)}_j(u;\{z_i\},\{v_i\})|j=1,\ldots,M\}$
\bea
H^{(I)}_j(u;\{z_i\},\{v_i\})&=&F_1(u)\prod_{l=1}^N\frac{\sin(u+z_l)}{\sin(u+z_l+\eta)}
      \frac{\prod_{k\neq j}\sin(u+v_k+2\eta)\sin(u-v_k+\eta)}
      {\sin(u-v_j)\sin(u+v_j+\eta)\sin(2u+\eta)}\no\\
      &&-F_2(u)\hspace{-0.1truecm}\prod_{l=1}^N\hspace{-0.1truecm}
      \frac{\sin(u-z_l+\eta)}{\sin(u-z_l)}
      \frac{\prod_{k\neq j}\sin(u+v_k)\sin(u-v_k-\eta)}
      {\sin(u\hspace{-0.1truecm}-\hspace{-0.1truecm}v_j)
      \sin(u\hspace{-0.1truecm}+\hspace{-0.1truecm}v_j
      \hspace{-0.1truecm}+\hspace{-0.1truecm}\eta)
      \sin(2u\hspace{-0.1truecm}+\hspace{-0.1truecm}\eta)},\\
H^{(II)}_j(u;\{z_i\},\{v_i\})&=&F_3(u)\prod_{l=1}^N\frac{\sin(u-z_l)}{\sin(u-z_l+\eta)}
      \frac{\prod_{k\neq j}\sin(v_k+u+2\eta)\sin(v_k-u-\eta)}
       {\sin(u+v_j+\eta)\sin(u-v_j)\sin(2u+\eta)}\no\\
      &&-F_4(u)\hspace{-0.1truecm}\prod_{l=1}^N\hspace{-0.1truecm}
      \frac{\sin(u+z_l+\eta)}{\sin(u+z_l)}
      \frac{\prod_{k\neq j}\sin(v_k+u)\sin(v_k-u+\eta)}
       {\sin(u\hspace{-0.1truecm}+\hspace{-0.1truecm}v_j
       \hspace{-0.1truecm}+\hspace{-0.1truecm}\eta)
       \sin(u\hspace{-0.1truecm}-\hspace{-0.1truecm}v_j)
       \sin(2u\hspace{-0.1truecm}+\hspace{-0.1truecm}\eta)},
\eea  where the coefficients $\{F_i(u)|i=1,2,3,4\}$ are
\bea
 F_1(u)&=& \sin(\l_2\hspace{-0.1truecm}+\hspace{-0.1truecm}\bar{\xi}
     \hspace{-0.1truecm}+\hspace{-0.1truecm}u
     \hspace{-0.1truecm}+\hspace{-0.1truecm}\eta)
     \sin(\l_2\hspace{-0.1truecm}+\hspace{-0.1truecm}\xi
     \hspace{-0.1truecm}-\hspace{-0.1truecm}u\hspace{-0.1truecm}-\hspace{-0.1truecm}\eta)
     \sin(\l_1\hspace{-0.1truecm}+\hspace{-0.1truecm}\bar{\xi}\hspace{-0.1truecm}-\hspace{-0.1truecm}
     u\hspace{-0.1truecm}-\hspace{-0.1truecm}\eta)
     \sin(\l_1\hspace{-0.1truecm}+\hspace{-0.1truecm}\xi\hspace{-0.1truecm}+\hspace{-0.1truecm}u
     \hspace{-0.1truecm}+\hspace{-0.1truecm}\eta),\\
 F_2(u)&=&\sin(\l_2+\bar{\xi}-u)\sin(\l_2+\xi+u)
     \sin(\l_1+\bar{\xi}+u)\sin(\l_1+\xi-u),\\
 F_3(u)&=& \sin(\l_2\hspace{-0.1truecm}+\hspace{-0.1truecm}\bar{\xi}
     \hspace{-0.1truecm}-\hspace{-0.1truecm}u\hspace{-0.1truecm}-\hspace{-0.1truecm}\eta)
     \sin(\l_2\hspace{-0.1truecm}+\hspace{-0.1truecm}\xi\hspace{-0.1truecm}+\hspace{-0.1truecm}u
     \hspace{-0.1truecm}+\hspace{-0.1truecm}\eta)
     \sin(\l_1\hspace{-0.1truecm}+\hspace{-0.1truecm}\bar{\xi}\hspace{-0.1truecm}+\hspace{-0.1truecm}u\hspace{-0.1truecm}
     +\hspace{-0.1truecm}\eta)
     \sin(\l_1\hspace{-0.1truecm}+\hspace{-0.1truecm}\xi
     \hspace{-0.1truecm}-\hspace{-0.1truecm}u\hspace{-0.1truecm}-\hspace{-0.1truecm}\eta),\\
 F_4(u)&=&\sin(\l_2+\bar{\xi}+u)\sin(\l_2+\xi-u)
     \sin(\l_1+\bar{\xi}-u)\sin(\l_1+\xi+u).
\eea

Let us consider the scalar product $S^{I,I}(\{u_{\a}\};\{v^{(1)}_i\})$ defined by (\ref{Scalar-2}). The expression (\ref{Scalar-6})
of $S^{I,I}(\{u_{\a}\};\{v^{(1)}_i\})$ under the F-basis and the polarization free representations
(\ref{Creation-operator-1}) and (\ref{Creation-operator-2}) of the
pseudo-particle creation/annihilation operators allow us to compute the scalar product following the similar procedure as that in
\cite{Kit99} for the bulk case as follows. In front of each operators $\tilde{\T}^-_F$ in (\ref{Scalar-6}), we insert
a sum over the complete set of spin states $|j_1,\ldots,j_i\gg$, where $|j_1,\ldots,j_i\gg$ is the state with $i$ spins
being $\e_2$  in the sites $j_1,\ldots,j_i$ and $2M-i$ spins being $\e_1$ in the other sites. We are thus led to consider
some intermediate functions of the form
\bea
 G^{(i)}(u_1,\ldots,u_i|j_{i+1},\ldots,j_M;\{v^{(1)}_i\})
    \hspace{-0.38truecm}&=&\hspace{-0.38truecm}
    \ll \hspace{-0.1truecm} j_{i+1},\ldots,j_M \hspace{-0.1truecm}|
    \tilde{\T}^-_F(\l\hspace{-0.1truecm}-\hspace{-0.1truecm}2(i\hspace{-0.1truecm}-\hspace{-0.1truecm}1)
    \eta\hat{1},\l|u_i)^2_1\ldots \tilde{\T}^-_F(\l,\l|u_1)^2_1\no\\
 \hspace{-0.38truecm}&&\hspace{-0.38truecm}\times \tilde{\T}^+_F(\l,\l
    \hspace{-0.1truecm}+\hspace{-0.1truecm}2\eta\hat{1}|v^{(1)}_1)^1_2 \ldots
    \tilde{\T}^+_F(\l,\l\hspace{-0.1truecm}+\hspace{-0.1truecm}2M\eta\hat{1}|v^{(1)}_M)^1_2|1,\ldots,1\rangle,\no\\
 &&\qquad\qquad\qquad\qquad\qquad\qquad i=0,1,\ldots,M,
\eea which satisfy the following recursive relation:
\bea
  &&G^{(i)}(u_1,\ldots,u_i|j_{i+1},\ldots,j_M;\{v^{(1)}_i\})\no\\
  &&\qquad\qquad=\sum_{j\neq j_{i+1},\ldots,j_M}\ll \hspace{-0.1truecm} j_{i+1},\ldots,j_M \hspace{-0.1truecm}|
    \tilde{\T}^-_F(\l\hspace{-0.1truecm}-\hspace{-0.1truecm}2(i\hspace{-0.1truecm}-\hspace{-0.1truecm}1)
    \eta\hat{1},\l|u_i)^2_1|j,j_{i+1},\ldots,j_M\hspace{-0.1truecm}\gg\no\\
  &&\quad\quad\quad\qquad\qquad\times G^{(i-1)}(u_1,\ldots,u_{i-1}|j,j_{i+1},\ldots,j_M;\{v^{(1)}_i\}),
  \quad i=1,\ldots,M.\label{Recursive}
\eea Note that the last of these functions $\{G^{(i)}|i=0,\ldots,M\}$ is precisely the scalar product $S^{I,I}(\{u_{\a}\};\{v^{(1)}_i\})$,
namely,
\bea
 G^{(M)}(u_1,\ldots,u_M;\{v^{(1)}_i\})=S^{I,I}(\{u_{\a}\};\{v^{(1)}_i\}),
\eea whereas the first one,
\bea
 G^{(0)}(j_{1},\ldots,j_M;\{v^{(1)}_i\})=\ll \hspace{-0.1truecm} j_{1},\ldots,j_M \hspace{-0.1truecm}|
  \tilde{\T}^+_F(\l,\l\hspace{-0.1truecm}+\hspace{-0.1truecm}2\eta\hat{1}|v^{(1)}_1)^1_2
  \ldots\tilde{\T}^+_F(\l,\l\hspace{-0.1truecm}+\hspace{-0.1truecm}2M\eta\hat{1}|v^{(1)}_M)^1_2
  |1,\ldots,1\rangle,\no
\eea  is closely related to the partition function computed in \cite{Yan10}. Solving the recursive relations
(\ref{Recursive}), we find that
if the parameters $\{v^{(1)}_k\}$ satisfy the first set of  Bethe
ansatz equations (\ref{BA-D-1}) the scalar product
$S^{I,I}(\{u_{\a}\}; \{v^{(1)}_i\})$ has the following determinant representation
\bea
S^{I,I}(\{u_{\a}\};\{v^{(1)}_i\})
  \hspace{-0.38truecm}&=&\hspace{-0.4truecm}\prod_{k=1}^M\hspace{-0.1truecm}\lt\{\hspace{-0.1truecm}
  \frac{\sin(\l_{12}\hspace{-0.1truecm}+\hspace{-0.1truecm}2\eta\hspace{-0.1truecm}-\hspace{-0.1truecm}2k\eta)
  \sin(\l_{12}\hspace{-0.1truecm}-\hspace{-0.1truecm}\eta\hspace{-0.1truecm}+\hspace{-0.1truecm}2k\eta)}
  {\sin(\l_{12}\hspace{-0.1truecm}-\hspace{-0.1truecm}
  (k\hspace{-0.1truecm}-\hspace{-0.1truecm}1)\eta)
  \sin(\l_{12}\hspace{-0.1truecm}+\hspace{-0.1truecm}k\eta)}\hspace{-0.1truecm}
  \prod_{l=1}^{N}\hspace{-0.1truecm}\frac{\sin(u_k-z_l)\sin(v^{(1)}_k-z_l)}
  {\sin(u_k\hspace{-0.1truecm}-\hspace{-0.1truecm}z_l\hspace{-0.1truecm}+\hspace{-0.1truecm}\eta)
  \sin(v^{(1)}_k\hspace{-0.1truecm}-\hspace{-0.1truecm}z_l\hspace{-0.1truecm}+\hspace{-0.1truecm}\eta)}
  \hspace{-0.1truecm}\rt\}\no\\[6pt]
\hspace{-0.38truecm}&&\hspace{-0.28truecm}\times\hspace{-0.1truecm}
  \frac{{\rm det}\bar{{\cal N}}^{(I)}(\{u_{\a}\};\{v^{(1)}_i\})}
  {\prod_{\a<\b}\sin(u_{\a}
  \hspace{-0.1truecm}-\hspace{-0.1truecm}u_{\b})
  \sin(u_{\a}\hspace{-0.1truecm}+\hspace{-0.1truecm}u_{\b}
  \hspace{-0.1truecm}+\hspace{-0.1truecm}\eta)
  \hspace{-0.1truecm}\prod_{k>l}\sin(v^{(1)}_k
  \hspace{-0.1truecm}-\hspace{-0.1truecm}v^{(1)}_l)\sin(v^{(1)}_k
  \hspace{-0.1truecm}+\hspace{-0.1truecm}v^{(1)}_l
  \hspace{-0.1truecm}+\hspace{-0.1truecm}\eta)},\no\\
  \label{Determinant-1}
\eea where the $M\times M$ matrix $\bar{{\cal N}}^{(I)}(\{u_{\a}\};\{v^{(1)}_i\})$ is given by
\bea
 \bar{{\cal N}}^{(I)}(\{u_{\a}\};\{v^{(1)}_i\})_{\a,j}\hspace{-0.1truecm}=\hspace{-0.1truecm}\frac
 {\sin\eta\sin(2u_{\a})\sin(2v^{(1)}_j+2\eta)H^{(I)}_j(u_{\a};\{z_i\},\{v^{(1)}_i\})}
 {\sin(\l_1\hspace{-0.1truecm}+\hspace{-0.1truecm}\xi
 \hspace{-0.1truecm}+\hspace{-0.1truecm}u_{\a})
 \sin(\l_2\hspace{-0.1truecm}+\hspace{-0.1truecm}\xi\hspace{-0.1truecm}+\hspace{-0.1truecm}u_{\a})
 \sin(\l_2\hspace{-0.1truecm}+\hspace{-0.1truecm}\bar{\xi}
 \hspace{-0.1truecm}-\hspace{-0.1truecm}v^{(1)}_j
 \hspace{-0.1truecm}-\hspace{-0.1truecm}\eta)\sin(\l_1\hspace{-0.1truecm}+
 \hspace{-0.1truecm}\bar{\xi}\hspace{-0.1truecm}-\hspace{-0.1truecm}v^{(1)}_{j}
 \hspace{-0.1truecm}-\hspace{-0.1truecm}\eta)}.\no\\
\eea Using the similar method as above, we have that the scalar product
$S^{II,II}(\{u_{\a}\}; \{v^{(2)}_i\})$ has the following determinant representation provided that
the parameters  $\{v^{(2)}_k\}$ satisfy the second set of  Bethe
ansatz equations (\ref{BA-D-2})
\bea
S^{II,II}(\{u_{\a}\};\{v^{(2)}_i\})
  \hspace{-0.38truecm}&=&\hspace{-0.4truecm}\prod_{k=1}^M\hspace{-0.1truecm}\lt\{\hspace{-0.1truecm}
  \frac{\sin(\l_{12}\hspace{-0.1truecm}+\hspace{-0.1truecm}2k\eta)
  \sin(\l_{21}\hspace{-0.1truecm}-\hspace{-0.1truecm}\eta
  \hspace{-0.1truecm}+\hspace{-0.1truecm}2k\eta)}
  {\sin(\l_{12}\hspace{-0.1truecm}+\hspace{-0.1truecm}k\eta)
  \sin(\l_{21}\hspace{-0.1truecm}+\hspace{-0.1truecm}(k\hspace{-0.1truecm}-\hspace{-0.1truecm}1)\eta)}\hspace{-0.1truecm}
  \prod_{l=1}^{N}\hspace{-0.1truecm}\frac{\sin(u_k+z_l)\sin(v^{(2)}_k+z_l)}
  {\sin(u_k\hspace{-0.1truecm}+\hspace{-0.1truecm}z_l\hspace{-0.1truecm}+\hspace{-0.1truecm}\eta)
  \sin(v^{(2)}_k\hspace{-0.1truecm}+\hspace{-0.1truecm}z_l\hspace{-0.1truecm}+\hspace{-0.1truecm}\eta)}
  \hspace{-0.1truecm}\rt\}\no\\[6pt]
\hspace{-0.38truecm}&&\hspace{-0.28truecm}\times\hspace{-0.1truecm}
  \frac{{\rm det}\bar{{\cal N}}^{(II)}(\{u_{\a}\};\{v^{(2)}_i\})}
  {\prod_{\a<\b}\sin(u_{\a}
  \hspace{-0.1truecm}-\hspace{-0.1truecm}u_{\b})
  \sin(u_{\a}\hspace{-0.1truecm}+\hspace{-0.1truecm}u_{\b}
  \hspace{-0.1truecm}+\hspace{-0.1truecm}\eta)
  \hspace{-0.1truecm}\prod_{k>l}\sin(v^{(2)}_k
  \hspace{-0.1truecm}-\hspace{-0.1truecm}v^{(2)}_l)\sin(v^{(2)}_k
  \hspace{-0.1truecm}+\hspace{-0.1truecm}v^{(2)}_l
  \hspace{-0.1truecm}+\hspace{-0.1truecm}\eta)},\no\\
  \label{Determinant-2}
\eea where the $M\times M$ matrix $\bar{{\cal N}}^{(II)}(\{u_{\a}\};\{v^{(2)}_i\})$ is given by
\bea
 \bar{{\cal N}}^{(II)}(\{u_{\a}\};\{v^{(2)}_i\})_{\a,j}\hspace{-0.1truecm}=\hspace{-0.1truecm}\frac
 {\sin\eta\sin(2u_{\a}+2\eta)\sin(2v^{(2)}_j)H^{(II)}_j(u_{\a};\{z_i\},\{v^{(2)}_i\})}
 {\sin(\l_2\hspace{-0.1truecm}+\hspace{-0.1truecm}\bar{\xi}
 \hspace{-0.1truecm}-\hspace{-0.1truecm}u_{\a}\hspace{-0.1truecm}-\hspace{-0.1truecm}\eta)
 \sin(\l_1\hspace{-0.1truecm}+\hspace{-0.1truecm}\bar{\xi}\hspace{-0.1truecm}-\hspace{-0.1truecm}u_{\a}
 \hspace{-0.1truecm}-\hspace{-0.1truecm}\eta)
 \sin(\l_2\hspace{-0.1truecm}+\hspace{-0.1truecm}\xi
 \hspace{-0.1truecm}+\hspace{-0.1truecm}v^{(2)}_j)\sin(\l_1\hspace{-0.1truecm}+
 \hspace{-0.1truecm}\xi\hspace{-0.1truecm}+\hspace{-0.1truecm}v^{(2)}_{j})}.\no\\
\eea

Now we are in position to compute the norms of the Bethe states which can be obtained
by taking the limit $u_{\a}\rightarrow v^{(i)}_{\a},$ $\a=1,\ldots M$. The norm of the first set of
Bethe state (\ref{Bethe-state-1}) is
\bea
 \mathbb{N}^{I,I}(\{v^{(1)}_{\a}\})&=&\lim_{u_{\a}\rightarrow v^{(1)}_{\a}}S^{I,I}(\{u_{\a}\}; \{v^{(1)}_i\})\no\\[6pt]
   &=&\prod_{k=1}^M\hspace{-0.1truecm}\lt\{\hspace{-0.1truecm}
  \frac{\sin(\l_{12}\hspace{-0.1truecm}+\hspace{-0.1truecm}2\eta\hspace{-0.1truecm}-\hspace{-0.1truecm}2k\eta)
  \sin(\l_{12}\hspace{-0.1truecm}-\hspace{-0.1truecm}\eta\hspace{-0.1truecm}+\hspace{-0.1truecm}2k\eta)}
  {\sin(\l_{12}\hspace{-0.1truecm}-\hspace{-0.1truecm}
  (k\hspace{-0.1truecm}-\hspace{-0.1truecm}1)\eta)
  \sin(\l_{12}\hspace{-0.1truecm}+\hspace{-0.1truecm}k\eta)}\hspace{-0.1truecm}
  \prod_{l=1}^{N}\hspace{-0.1truecm}\frac{\sin^2(v^{(1)}_k-z_l)}
  {\sin^2(v^{(1)}_k\hspace{-0.1truecm}-\hspace{-0.1truecm}z_l\hspace{-0.1truecm}+\hspace{-0.1truecm}\eta)}
  \hspace{-0.1truecm}\rt\}\no\\[6pt]
  &&\times \prod_{\a\ne \b}\frac{\sin(v^{(1)}_{\a}+v^{(1)}_{\b})\sin(v^{(1)}_{\a}-v^{(1)}_{\b}-\eta)}
  {\sin(v^{(1)}_{\a}-v^{(1)}_{\b})\sin(v^{(1)}_{\a}+v^{(1)}_{\b}+\eta)} \, {\rm det}\Phi^{(I)}(\{v^{(1)}_{\a}\}),
  \label{Norm-1}
\eea where the matrix elements of $M\times M$ matrix $\Phi^{(I)}(\{v_{\a}\})$ are given by
\bea
  \Phi^{(I)}_{\a,j}(\{v_{\a}\})&=& \frac{\sin\eta
     \sin(\l_2\hspace{-0.1truecm}+\hspace{-0.1truecm}\bar{\xi}\hspace{-0.1truecm}-\hspace{-0.1truecm}v_{\a})
     \sin(\l_1\hspace{-0.1truecm}+\hspace{-0.1truecm}\bar{\xi}\hspace{-0.1truecm}+\hspace{-0.1truecm}v_{\a})
     \sin(\l_1\hspace{-0.1truecm}+\hspace{-0.1truecm}\xi-\hspace{-0.1truecm}v_{\a})}
     {\sin(\l_1\hspace{-0.1truecm}+\hspace{-0.1truecm}\xi\hspace{-0.1truecm}+\hspace{-0.1truecm}v_{\a})
     \sin(\l_2\hspace{-0.1truecm}+\hspace{-0.1truecm}\bar{\xi}\hspace{-0.1truecm}-\hspace{-0.1truecm}v_j\hspace{-0.1truecm}-\hspace{-0.1truecm}\eta)
     \sin(\l_1\hspace{-0.1truecm}+\hspace{-0.1truecm}\bar{\xi}\hspace{-0.1truecm}-\hspace{-0.1truecm}v_j\hspace{-0.1truecm}-\hspace{-0.1truecm}\eta)}
     \no\\[6pt]
  &&\times \frac{\sin2v_{\a}\sin(2v_j\hspace{-0.1truecm}+\hspace{-0.1truecm}2\eta)}
     {\sin(2v_{\a}\hspace{-0.1truecm}+\hspace{-0.1truecm}\eta)\sin(2v_j\hspace{-0.1truecm}+\hspace{-0.1truecm}\eta)}
     \prod_{l=1}^N\frac{\sin(v_{\a}-z_l+\eta)}{\sin(v_{\a}-z_l)}\no\\[6pt]
  &&\times\frac{\partial}{\partial v_{\a}}\,\ln\lt\{
     \frac{\sin(\l_2\hspace{-0.1truecm}+\hspace{-0.1truecm}\bar{\xi}\hspace{-0.1truecm}+\hspace{-0.1truecm}v_j\hspace{-0.1truecm}+\hspace{-0.1truecm}\eta)
     \sin(\l_2\hspace{-0.1truecm}+\hspace{-0.1truecm}\xi\hspace{-0.1truecm}-\hspace{-0.1truecm}v_j
     \hspace{-0.1truecm}-\hspace{-0.1truecm}\eta)
     \sin(\l_1\hspace{-0.1truecm}+\hspace{-0.1truecm}\bar{\xi}\hspace{-0.1truecm}-\hspace{-0.1truecm}v_j
     \hspace{-0.1truecm}-\hspace{-0.1truecm}\eta)
     \sin(\l_1\hspace{-0.1truecm}+\hspace{-0.1truecm}\xi+\hspace{-0.1truecm}v_j\hspace{-0.1truecm}+\hspace{-0.1truecm}\eta)}
     {\sin(\l_2\hspace{-0.1truecm}+\hspace{-0.1truecm}\bar{\xi}\hspace{-0.1truecm}-\hspace{-0.1truecm}v_j)
     \sin(\l_2\hspace{-0.1truecm}+\hspace{-0.1truecm}\xi\hspace{-0.1truecm}+\hspace{-0.1truecm}v_j)
     \sin(\l_1\hspace{-0.1truecm}+\hspace{-0.1truecm}\bar{\xi}\hspace{-0.1truecm}+\hspace{-0.1truecm}v_j)
     \sin(\l_1\hspace{-0.1truecm}+\hspace{-0.1truecm}\xi\hspace{-0.1truecm}-\hspace{-0.1truecm}v_j)}\rt.\no\\[6pt]
  &&\quad\quad\times \prod_{l=1}^N\lt.
     \frac{\sin(v_j\hspace{-0.1truecm}+\hspace{-0.1truecm}z_l)\sin(v_j\hspace{-0.1truecm}-\hspace{-0.1truecm}z_l)}
     {\sin(v_j\hspace{-0.1truecm}+\hspace{-0.1truecm}z_l\hspace{-0.1truecm}+\hspace{-0.1truecm}\eta)
     \sin(v_j\hspace{-0.1truecm}-\hspace{-0.1truecm}z_l\hspace{-0.1truecm}+\hspace{-0.1truecm}\eta)}
     \prod_{k\ne j}
     \frac{\sin(v_j\hspace{-0.1truecm}+\hspace{-0.1truecm}v_k\hspace{-0.1truecm}+\hspace{-0.1truecm}2\eta)
     \sin(v_j\hspace{-0.1truecm}-\hspace{-0.1truecm}v_k\hspace{-0.1truecm}+\hspace{-0.1truecm}\eta)}
     {\sin(v_j+v_k)\sin(v_j\hspace{-0.1truecm}-\hspace{-0.1truecm}v_k-\eta)}\rt\},
     \label{Norm-2}
\eea the norm of the second set of
Bethe state (\ref{Bethe-state-2}) is given by
\bea
 \mathbb{N}^{II,II}(\{v^{(2)}_{\a}\})&=&\lim_{u_{\a}\rightarrow v^{(2)}_{\a}}S^{II,II}(\{u_{\a}\}; \{v^{(2)}_i\})\no\\[6pt]
   &=&\prod_{k=1}^M\hspace{-0.1truecm}\lt\{\hspace{-0.1truecm}
  \frac{\sin(\l_{12}\hspace{-0.1truecm}+\hspace{-0.1truecm}2k\eta)
  \sin(\l_{21}\hspace{-0.1truecm}-\hspace{-0.1truecm}\eta\hspace{-0.1truecm}+\hspace{-0.1truecm}2k\eta)}
  {\sin(\l_{12}\hspace{-0.1truecm}+\hspace{-0.1truecm}k\eta)
  \sin(\l_{21}\hspace{-0.1truecm}+\hspace{-0.1truecm}(k\hspace{-0.1truecm}-\hspace{-0.1truecm}1)\eta)}\hspace{-0.1truecm}
  \prod_{l=1}^{N}\hspace{-0.1truecm}\frac{\sin^2(v^{(2)}_k+z_l)}
  {\sin^2(v^{(2)}_k\hspace{-0.1truecm}+\hspace{-0.1truecm}z_l\hspace{-0.1truecm}+\hspace{-0.1truecm}\eta)}
  \hspace{-0.1truecm}\rt\}\no\\[6pt]
  &&\times \prod_{\a\ne \b}\frac{\sin(v^{(2)}_{\a}\hspace{-0.1truecm}+\hspace{-0.1truecm}v^{(2)}_{\b}
  \hspace{-0.1truecm}+\hspace{-0.1truecm}2\eta)\sin(v^{(2)}_{\a}\hspace{-0.1truecm}-\hspace{-0.1truecm}v^{(2)}_{\b}
  \hspace{-0.1truecm}-\hspace{-0.1truecm}\eta)}
  {\sin(v^{(2)}_{\a}\hspace{-0.1truecm}-\hspace{-0.1truecm}v^{(2)}_{\b})\sin(v^{(2)}_{\a}
  \hspace{-0.1truecm}+\hspace{-0.1truecm}v^{(2)}_{\b}\hspace{-0.1truecm}+\hspace{-0.1truecm}\eta)} \, {\rm det}\Phi^{(II)}(\{v^{(2)}_{\a}\}),
  \label{Norm-3}
\eea where the matrix elements of $M\times M$ matrix $\Phi^{(II)}(\{v_{\a}\})$ are given by
\bea
  \Phi^{(II)}_{\a,j}(\{v_{\a}\})&=& \frac{\sin\eta
     \sin(\l_2\hspace{-0.1truecm}+\hspace{-0.1truecm}\xi\hspace{-0.1truecm}+\hspace{-0.1truecm}v_{\a}\hspace{-0.1truecm}+\hspace{-0.1truecm}\eta)
     \sin(\l_1\hspace{-0.1truecm}+\hspace{-0.1truecm}\bar{\xi}\hspace{-0.1truecm}+\hspace{-0.1truecm}v_{\a}\hspace{-0.1truecm}+\hspace{-0.1truecm}\eta)
     \sin(\l_1\hspace{-0.1truecm}+\hspace{-0.1truecm}\xi-\hspace{-0.1truecm}v_{\a}\hspace{-0.1truecm}-\hspace{-0.1truecm}\eta)}
     {\sin(\l_1\hspace{-0.1truecm}+\hspace{-0.1truecm}\bar{\xi}\hspace{-0.1truecm}-\hspace{-0.1truecm}v_{\a}\hspace{-0.1truecm}-\hspace{-0.1truecm}\eta)
     \sin(\l_2\hspace{-0.1truecm}+\hspace{-0.1truecm}\xi\hspace{-0.1truecm}+\hspace{-0.1truecm}v_j)
     \sin(\l_1\hspace{-0.1truecm}+\hspace{-0.1truecm}\xi\hspace{-0.1truecm}+\hspace{-0.1truecm}v_j)}
     \no\\[6pt]
  &&\times \frac{\sin(2v_{\a}\hspace{-0.1truecm}+\hspace{-0.1truecm}2\eta)\sin2v_j}
     {\sin(2v_{\a}\hspace{-0.1truecm}+\hspace{-0.1truecm}\eta)\sin(2v_j\hspace{-0.1truecm}+\hspace{-0.1truecm}\eta)}
     \prod_{l=1}^N\frac{\sin(v_{\a}-z_l)}{\sin(v_{\a}-z_l+\eta)}\no\\[6pt]
  &&\times\frac{\partial}{\partial v_{\a}}\,\ln\lt\{
     \frac{\sin(\l_2\hspace{-0.1truecm}+\hspace{-0.1truecm}\bar{\xi}\hspace{-0.1truecm}-\hspace{-0.1truecm}v_j\hspace{-0.1truecm}-\hspace{-0.1truecm}\eta)
     \sin(\l_2\hspace{-0.1truecm}+\hspace{-0.1truecm}\xi\hspace{-0.1truecm}+\hspace{-0.1truecm}v_j
     \hspace{-0.1truecm}+\hspace{-0.1truecm}\eta)
     \sin(\l_1\hspace{-0.1truecm}+\hspace{-0.1truecm}\bar{\xi}\hspace{-0.1truecm}+\hspace{-0.1truecm}v_j
     \hspace{-0.1truecm}+\hspace{-0.1truecm}\eta)
     \sin(\l_1\hspace{-0.1truecm}+\hspace{-0.1truecm}\xi-\hspace{-0.1truecm}v_j\hspace{-0.1truecm}-\hspace{-0.1truecm}\eta)}
     {\sin(\l_2\hspace{-0.1truecm}+\hspace{-0.1truecm}\bar{\xi}\hspace{-0.1truecm}+\hspace{-0.1truecm}v_j)
     \sin(\l_2\hspace{-0.1truecm}+\hspace{-0.1truecm}\xi\hspace{-0.1truecm}-\hspace{-0.1truecm}v_j)
     \sin(\l_1\hspace{-0.1truecm}+\hspace{-0.1truecm}\bar{\xi}\hspace{-0.1truecm}-\hspace{-0.1truecm}v_j)
     \sin(\l_1\hspace{-0.1truecm}+\hspace{-0.1truecm}\xi\hspace{-0.1truecm}+\hspace{-0.1truecm}v_j)}\rt.\no\\[6pt]
  &&\quad\quad\times \prod_{l=1}^N\lt.
     \frac{\sin(v_j\hspace{-0.1truecm}+\hspace{-0.1truecm}z_l)\sin(v_j\hspace{-0.1truecm}-\hspace{-0.1truecm}z_l)}
     {\sin(v_j\hspace{-0.1truecm}+\hspace{-0.1truecm}z_l\hspace{-0.1truecm}+\hspace{-0.1truecm}\eta)
     \sin(v_j\hspace{-0.1truecm}-\hspace{-0.1truecm}z_l\hspace{-0.1truecm}+\hspace{-0.1truecm}\eta)}
     \prod_{k\ne j}
     \frac{\sin(v_j\hspace{-0.1truecm}+\hspace{-0.1truecm}v_k\hspace{-0.1truecm}+\hspace{-0.1truecm}2\eta)
     \sin(v_j\hspace{-0.1truecm}-\hspace{-0.1truecm}v_k\hspace{-0.1truecm}+\hspace{-0.1truecm}\eta)}
     {\sin(v_j+v_k)\sin(v_j\hspace{-0.1truecm}-\hspace{-0.1truecm}v_k-\eta)}\rt\}.
     \label{Norm-4}
\eea Moreover, one may check that if the parameters $\{u_{\a}\}$ satisfy the Bethe ansatz equations (i.e. on shell) but different
from $\{v^{(i)}_{\a}\}$ the corresponding scalar products $S^{I,I}(\{u_{\a}\}; \{v^{(1)}_i\})$ or $S^{II,II}(\{u_{\a}\}; \{v^{(2)}_i\})$
vanishes, namely, the corresponding Bethe states are orthogonal.

%%%%%%%%%%%%%%%%%%%%%%%%%%%%%%%%%%%%%%%%%%%%%%%%%%%%%%%%%%%%%%%
%                                                             %
%  6. Conclusions                                             %
%                                                             %
%                                                             %
%                                                             %
%%%%%%%%%%%%%%%%%%%%%%%%%%%%%%%%%%%%%%%%%%%%%%%%%%%%%%%%%%%%%%%

\section{ Conclusions}
\label{C} \setcounter{equation}{0}

We have studied scalar products between an
on-shell Bethe state and  a general state (or an off-shell Bethe state) of
the open XXZ chain with non-diagonal boundary
terms, where the non-diagonal K-matrices $K^{\pm}(u)$ are given by
(\ref{K-matrix-2-1}) and (\ref{K-matrix-6}).  In our calculation
the factorizing F-matrix (\ref{F-matrix}) in the face picture of the
open XXZ chain, which leads to the polarization free representations
(\ref{Creation-operator-1}) and (\ref{Creation-operator-2}) of the associated
pseudo-particle creation/annihilation  operators, has played an important role. It is found that  the scalar
products can be expressed in terms of the determinants (\ref{partition-1}), (\ref{partition-2}),
(\ref{Determinant-1}) and (\ref{Determinant-2}). By taking the on shell limit, we
obtain the determinant representations (or Gaudin formula) (\ref{Norm-1})-(\ref{Norm-2}) and (\ref{Norm-3})-(\ref{Norm-4})
of the norms of the Bethe states.

%%%%%%%%%%%%%%%%%%%%%%%%%%%%%%%%%%%%%%%%%%%%%%%%%%%%%%%%%%%%%%%
%                                                             %
%  Acknowledgments                                            %
%                                                             %
%%%%%%%%%%%%%%%%%%%%%%%%%%%%%%%%%%%%%%%%%%%%%%%%%%%%%%%%%%%%%%%
\section*{Acknowledgements}
The financial supports from  the National Natural Science
Foundation of China (Grant No. 11075126 and 11031005), Australian Research Council
and the NWU Graduate Cross-discipline Fund (08YJC24)
are gratefully acknowledged. One of authors (W.L.Y.) would like to thank
the school of Mathematics and Statistics of the University of Sydney where
part of this work was done,
especially Xin Liu, for hospitality.

%%%%%%%%%%%%%%%%%%%%%%%%%%%%%%%%%%%%%%%%%%%%%%%%%%%%%%%%%%%%%%%
%                                                             %
%  Appendix A                                                 %
%                                                             %
%%%%%%%%%%%%%%%%%%%%%%%%%%%%%%%%%%%%%%%%%%%%%%%%%%%%%%%%%%%%%%%

\section*{Appendix A: $\T^{\pm}(m|u)$ in the face picture }
\setcounter{equation}{0}
\renewcommand{\theequation}{A.\arabic{equation}}

The K-matrices $K^{\pm}(u)$ given by (\ref{K-matrix}) and
(\ref{DK-matrix}) are generally non-diagonal (in the vertex
picture), after the face-vertex transformations (\ref{K-F-1}) and
(\ref{K-F-2}), the face type counterparts $\K(\l|u)$ and
$\tilde{\K}(\l|u)$ given by (\ref{K-F-3}) and (\ref{K-F-4}) {\it
simultaneously\/} become diagonal. This fact suggests that it
would be much simpler if one performs all calculations in the face
picture.

Associated with the vertex type monodromy matrices $T(u)$
(\ref{Mon-V}) and $\hat{T}(u)$ (\ref{Mon-V-0}), we introduce the
following operators \bea
 T(m,l|u)^j_{\mu}&=&\tilde{\phi}^0_{m+\eta\hat{\jmath},m}(u)\,T_0(u)\,
    \phi^0_{l+\eta\hat{\mu},l}(u),\\
 S(m,l|u)^{\mu}_{i}&=&\bar{\phi}^0_{l,l-\eta\hat{\mu}}(-u)\,\hat{T}_0(u)\,
    \phi^0_{m,m-\eta\hat{\imath}}(-u).
\eea Moreover, for the case of
$m=l-\eta\sum_{k=1}^N\hat{\imath}_k$, we introduce a generic state
in the quantum space from  the intertwiner vector (\ref{Intvect})
\bea
 |i_1,\ldots,i_N\rangle^{m}_{l}=
     \phi^1_{l,l-\eta\hat{\imath}_1}(z_1)
     \phi^2_{l-\eta\hat{\imath}_1,l-\eta(\hat{\imath}_1+\hat{\imath}_2)}(z_2)\ldots
     \phi^N_{l-\eta\sum_{k=1}^{N-1}\hat{\imath}_k,l-\eta\sum_{k=1}^{N}\hat{\imath}_k}(z_N).
     \label{Face-state}
\eea We can evaluate the action of the operator $T(m,l|u)$ on the
state $|i_1,\ldots,i_N\rangle^{m}_{l}$ from the face-vertex
correspondence relation (\ref{Face-vertex}) \bea
 &&T(m,l|u)^j_{\mu}|i_1,\ldots,i_N\rangle^{m}_{l}=
    \tilde{\phi}^0_{m+\eta\hat{\jmath},m}(u)\,T_0(u)\,
    \phi^0_{l+\eta\hat{\mu},l}(u)|i_1,\ldots,i_N\rangle^{m}_{l}
    \no\\
 &&\quad\quad=\tilde{\phi}^0_{m+\eta\hat{\jmath},m}(u)
    \R_{0,N}(u-z_N)\ldots\R_{0,1}(u-z_1)\phi^0_{l+\eta\hat{\mu},l}(u)
    \phi^1_{l,l-\eta\hat{\imath}_1}(z_1)\ldots\no\\
 &&\quad\quad=\sum_{\a_1,i'_1}R(u-z_1;l+\eta\hat{\mu})^{\a_1i'_1}_{\mu\,\, i_1}
   \phi^1_{l+\eta\hat{\mu},l+\eta\hat{\mu}-\eta\hat{\imath}'_1}(z_1)
   \tilde{\phi}^0_{m+\eta\hat{\jmath},m}(u)
    \R_{0,N}(u-z_N)\ldots\no\\
 &&\quad\quad\quad\quad \times
   \R_{0,2}(u-z_2)\phi^0_{l+\eta\hat{\mu}-\eta\hat{\imath}'_1,l-\eta\hat{\imath}_1}(u)
    \phi^2_{l-\eta\hat{\imath}_1,l-\eta(\hat{\imath}_1+\hat{\imath}_2)}(z_2)\ldots\no\\
 &&\quad\quad\vdots\no\\
 &&\quad\quad=\sum_{\a_{1}\ldots\a_{N-1}}\sum_{i'_1\ldots i'_N}
    R(u-z_N;l+\eta\hat{\mu}-\eta\sum_{k=1}^{N-1}\hat{\imath}'_k)
    ^{j\,\,\,\,\,\,\,\,\,\,\,\,i'_N}_{\a_{N-1}i_N}\ldots\no\\
 &&\quad\quad\quad\quad\times R(u-z_1;l+\eta\hat{\mu})^{\a_1i'_1}_{\mu \,\,i_1}
    |i'_1,\ldots,i'_N\rangle^{l+\eta\hat{\mu}-\eta\sum_{k=1}^N\hat{\imath}'_k}_{l+\eta\hat{\mu}}.
    \label{T-action}
\eea
Comparing with (\ref{Monodromy-face-2}), we have the following
correspondence \bea
 T(m,l|u)^j_{\mu}|i_1,\ldots,i_N\rangle^{m}_{l}\,\longleftrightarrow\,
      T_F(m+\eta\hat{\mu};l+\eta\hat{\mu}|u)_{\mu}^j|i_1,\ldots,i_N\rangle,\label{crospendence-1}
\eea where vector $|i_1,\ldots,i_N\rangle$ is given by
(\ref{Vector-V}). Hereafter, we will use  $O_F$ to denote the face
version of operator $O$ in the face picture.

Noting that \bea
 \hat{T}_0(u)=\R_{1,0}(u+z_1)\ldots\R_{N,0}(u+z_N),\no
\eea we obtain the action of $S(m,l|u)^{\mu}_{i}$ on the state
$|i_1,\ldots,i_N\rangle^{m}_{l}$ \bea
 S(m,l|u)^{\mu}_{i}|i_1,\ldots,i_N\rangle^{m}_{l}
     \hspace{-0.22truecm}&=&\hspace{-0.42truecm}
     \sum_{\a_{1}\ldots\a_{N-1}}\sum_{i'_1\ldots i'_N}
     R(u+z_1;l)^{i'_1\,\mu}_{i_1\a_{N-1}}
     R(u+z_2;l-\eta\hat{\imath}_1)^{i'_2\,\a_{N-1}}_{i_2\a_{N-2}}\no\\
 &&\,\times
     \ldots R(u\hspace{-0.12truecm}+\hspace{-0.12truecm}z_N;
     l\hspace{-0.12truecm}-\hspace{-0.12truecm}\eta
     \sum_{k=1}^{N-1}\hat{\imath}_k)
     ^{i'_N\,\a_1}_{i_N\,i}|i'_1,\ldots,i'_N\rangle
     ^{l-\eta\hat{\mu}-\eta\sum_{k=1}^N\hat{\imath}'_k}_{l-\eta\hat{\mu}}.\no\\
\eea Then the crossing relation of the R-matrix (\ref{Crossing})
enables us to establish the following relation:
\bea
  S(m,l|u)^{\mu}_{i}=\varepsilon_{\bar{i}}\varepsilon_{\bar{\mu}}
   \frac{\sin\lt(m_{21}\rt)}{\sin\lt(l_{21}\rt)}
   \prod_{k=1}^N\frac{\sin(u+z_k)}{\sin(u+z_k+\eta)}
   T(m,l|-u-\eta)^{\bar{i}}_{\bar{\mu}},
   \label{Crossing-operator}
\eea where the parities are defined in (\ref{Parity}) and $m_{21}$
(or $l_{21}$) is defined in (\ref{Def1}).

Now we are in the position to express $\T^{\pm}$ (\ref{Mon-F}) and
(\ref{Mon-F-1}) in terms of $T(m,l)^i_j$ and $S(l,m)^i_j$ which both can be expressed
in terms of the face type R-matrix (\ref{R-matrix}). By
(\ref{Int3}) and (\ref{Int4}),  we have \bea
  \T^-(m|u)^j_i&=&\tilde{\phi}^{0}_{m-\eta(\hat{\imath}-\hat{\jmath}),
      m-\eta\hat{\imath}}(u)~\mathbb{T}(u)~\phi^{0}_{m,
      m-\eta\hat{\imath}}(-u)\no\\
  &=&\tilde{\phi}^{0}_{m-\eta(\hat{\imath}-\hat{\jmath}),
      m-\eta\hat{\imath}}(u)T_0(u)K^-_0(u)\hat{T}_0(u)\phi^{0}_{m,
      m-\eta\hat{\imath}}(-u)\no\\
  &=&\hspace{-0.32truecm}\sum_{\mu,\nu}\tilde{\phi}^{0}_{m-\eta(\hat{\imath}-\hat{\jmath}),
      m-\eta\hat{\imath}}(u)T_0(u)
      \phi^0_{l-\eta(\hat{\nu}-\hat{\mu}),l-\eta\hat{\nu}}(u)
      \tilde{\phi}^0_{l-\eta(\hat{\nu}-\hat{\mu}),l-\eta\hat{\nu}}(u)\no\\
  &&\quad\times K^-_0(u)\phi^0_{l,l-\eta\hat{\nu}}(-u)
      \bar{\phi}^0_{l,l-\eta\hat{\nu}}(-u)
      \hat{T}_0(u)\phi^{0}_{m,
      m-\eta\hat{\imath}}(-u)\no\\
  &=&\hspace{-0.22truecm}\sum_{\mu,\nu}T(m-\eta\hat{\imath},l-\eta\hat{\nu}|u)^j_{\mu}
       \K(l|u)^{\mu}_{\nu}S(m,l|u)_i^{{\nu}}\no\\
  &\stackrel{{\rm def}}{=}& \T^-(m,l|u)^j_i,
\eea where the face-type K-matrix $\K(l|u)^{\mu}_{\nu}$ is given
by
\bea
 \K(l|u)^{\mu}_{\nu}=\tilde{\phi}^0_{l-\eta(\hat{\nu}-\hat{\mu}),l-\eta\hat{\nu}}(u)
    K^-_0(u)\phi^0_{l,l-\eta\hat{\nu}}(-u).\label{K-1}
\eea Similarly, we have
\bea
 \T^+(m|u)^j_i&=&\prod_{k\neq j}\frac{\sin m_{jk}}{\sin\lt(m_{jk}-\eta\rt)}
       \sum_{\mu,\nu}T(l-\eta\hat{\mu},m-\eta\hat{\jmath}|u)_i^{\nu}
       \tilde{\K}(l|u)^{\mu}_{\nu}S(l,m|u)^j_{{\mu}}\no\\
&\stackrel{{\rm
       def}}{=}&\T^+(l,m|u)^j_i
\eea with
\bea
 \tilde{\K}(l|u)^{\mu}_{\nu}=\bar{\phi}^0_{l,l-\eta\hat{\mu}}(-u)
    K^+_0(u)\phi^0_{l-\eta(\hat{\mu}-\hat{\nu}),l-\eta\hat{\mu}}(u).
    \label{K-2}
\eea Thanks to the fact that when $l=\l$ the corresponding
face-type K-matrices $\K(\l|u)$  (\ref{K-1}) and
$\tilde{\K}(\l|u)$  (\ref{K-2}) become diagonal ones (\ref{K-F-3})
and (\ref{K-F-4}), we have \bea
 \T^-(m,\l|u)^j_i\hspace{-0.22truecm}&=&\hspace{-0.22truecm}
       \sum_{\mu}T(m-\eta\hat{\imath},\l-\eta\hat{\mu}|u)^j_{\mu}
       k(\l|u)_{\mu}S(m,\l|u)_i^{{\mu}},\label{New-M-F-1}\\
 \T^+(\l,m|u)^j_i\hspace{-0.22truecm}&=&\hspace{-0.22truecm}
\prod_{k\neq j}\frac{\sin m_{jk}}{\sin\lt(m_{jk}-\eta\rt)}
       \sum_{\mu}T(\l-\eta\hat{\mu},m-\eta\hat{\jmath}|u)_i^{\mu}
       \tilde{k}(\l|u)_{\mu}S(\l,m|u)^j_{{\mu}},\label{New-M-F-2}
 \eea where the functions $k(\l|u)_{\mu}$ and $\tilde{k}(\l|u)_{\mu}$ are given by
 (\ref{K-F-3}) and (\ref{K-F-4}) respectively. The relation
(\ref{Crossing-operator}) implies that one can further express
$\T^{\pm}(m|u)^j_i$ in terms of only $T(m,l|u)^j_i$. Here we
present the results for the pseudo-particle creation operators
$\T^{-}(m|u)^2_1$ in (\ref{Bethe-state-2}) and $\T^{+}(m|u)^1_2$
in (\ref{Bethe-state-1}): \bea
 \T^-(m|u)^2_1\hspace{-0.22truecm}&=&\hspace{-0.22truecm}
      \T^-(m,\l|u)^2_1=\frac{\sin(m_{21})}{\sin(\l_{21})}\prod_{k=1}^N
      \frac{\sin(u+z_k)}{\sin(u+z_k+\eta)}\no\\
      &&\,\times\lt\{
      \frac{\sin(\l_1+\xi-u)}{\sin(\l_1+\xi+u)}
      T(m+\eta\hat{2},\l+\eta\hat{2}|u)^2_1
      T(m,\l|-u-\eta)^2_2\rt.\no\\
 &&\,\quad-\lt.
      \frac{\sin(\l_2+\xi-u)}{\sin(\l_2+\xi+u)}
      T(m+\eta\hat{2},\l+\eta\hat{1}|u)^2_2
      T(m,\l|-u-\eta)^2_1\rt\},\label{Expression-1}\\
 \T^+(m|u)^1_2\hspace{-0.22truecm}&=&\hspace{-0.22truecm}
      \T^+(\l,m|u)^1_2=\prod_{k=1}^N
      \frac{\sin(u+z_k)}{\sin(u+z_k+\eta)}\no\\
      &&\,\times\lt\{
      \frac{\sin(\l_{12}\hspace{-0.08truecm}-\hspace{-0.08truecm}\eta)
      \sin(\l_1\hspace{-0.08truecm}+\hspace{-0.08truecm}\bar{\xi}\hspace{-0.08truecm}+\hspace{-0.08truecm}u
      \hspace{-0.08truecm}+\hspace{-0.08truecm}\eta)}
      {\sin(m_{12}\hspace{-0.12truecm}-\hspace{-0.12truecm}\eta)
      \sin(\l_1\hspace{-0.12truecm}+\hspace{-0.12truecm}\bar{\xi}\hspace{-0.12truecm}-\hspace{-0.12truecm}u
      \hspace{-0.12truecm}-\hspace{-0.12truecm}\eta)}
      T(\l\hspace{-0.12truecm}+\hspace{-0.12truecm}\eta\hat{2},m\hspace{-0.12truecm}+\hspace{-0.12truecm}\eta\hat{2}|u)^1_2
      T(\l,m|\hspace{-0.12truecm}-\hspace{-0.12truecm}u\hspace{-0.12truecm}-\hspace{-0.12truecm}\eta)^2_2\rt.\no\\
 &&\,\quad\hspace{-0.12truecm}-\hspace{-0.12truecm}\lt.
      \frac{\sin(\l_{21}\hspace{-0.12truecm}-\hspace{-0.12truecm}\eta)
      \sin(\l_2\hspace{-0.12truecm}+\hspace{-0.12truecm}\bar{\xi}\hspace{-0.12truecm}+\hspace{-0.12truecm}u
      \hspace{-0.12truecm}+\hspace{-0.12truecm}\eta)}
      {\sin(m_{21}\hspace{-0.12truecm}+\hspace{-0.12truecm}\eta)
      \sin(\l_2\hspace{-0.12truecm}+\hspace{-0.12truecm}\bar{\xi}\hspace{-0.12truecm}-\hspace{-0.12truecm}u
      \hspace{-0.12truecm}-\hspace{-0.12truecm}\eta)}
      T(\l\hspace{-0.12truecm}+\hspace{-0.12truecm}\eta\hat{1},m
      \hspace{-0.12truecm}+\hspace{-0.12truecm}\eta\hat{2}|u)^2_2
      T(\l,m|\hspace{-0.12truecm}-\hspace{-0.12truecm}u\hspace{-0.12truecm}
      -\hspace{-0.12truecm}\eta)^1_2\rt\}.\no\\
 &&\label{Expression-2}
\eea Similar to (\ref{crospendence-1}), we have the
correspondence,
\bea
 &&\T^-(m,l|u)^2_1|i_1,\ldots,i_N\rangle^{m}_{l}\,\longleftrightarrow\,
   \T^-_{F}(m,l|u)^2_1|i_1,\ldots,i_N\rangle,\\
 &&\T^+(m,l|u)^1_2|i_1,\ldots,i_N\rangle^{m}_{l}\,\longleftrightarrow\,
   \T^+_{F}(m,l|u)^1_2|i_1,\ldots,i_N\rangle.
\eea This gives rise the expressions of the operators $\T^{\pm}_F(m|u)$ given by
(\ref{Expression-3}) and (\ref{Expression-4}).

Some remarks are in order. It follows from (\ref{T-action}) that the action of the operator
$T(m,l|u)$ on the state $|i_1,\ldots,i_N\rangle_l^m$ can be expressed in terms of the face type R-matrix
(\ref{R-matrix}). This implies that the corresponding actions of $\T^{\pm}(m|u)$ can also be
expressed in terms of the R-matrix and the K-matrices (\ref{K-F-3}) and (\ref{K-F-4}).
Moreover, the transfer
matrix $t(u)$ (\ref{trans}) can be given as a linear combination of
either $\T^{-}(m|u)^i_i$:
\bea
 \t(u)&=& tr(K^+(u)\mathbb{T}(u))=\sum_{\mu,\nu}tr\lt(K^+(u)\phi_{\l-\eta(\hat{\mu}-\hat{\nu}),\l-\eta\hat{\mu}}(u)
    \tilde{\phi}_{\l-\eta(\hat{\mu}-\hat{\nu}),\l-\eta\hat{\mu}}(u)\rt.\no\\
 &&\qquad\qquad \times \lt.\mathbb{T}(u)\phi_{\l,\l-\eta\hat{\mu}}(-u)\bar{\phi}_{\l,\l-\eta\hat{\mu}}(-u)\rt)\no\\
 &=&\sum_{\mu,\nu}\tilde{\K}(\l|u)_{\nu}^{\mu}\T^-(\l|u)^{\nu}_{\mu}=
\sum_{\mu}\tilde{k}(\l|u)_{\mu}\T^-(\l|u)^{\mu}_{\mu},
\eea or $\T^{+}(m|u)^i_i$:
\bea
 \t(u)&=& tr(K^+(u)\mathbb{T}(u))=tr\lt((\mathbb{T}^+(u))^{t_0}(K^-(u))^{t_0}\rt)\no\\
 &=&\sum_{\mu,\nu}tr\lt((\mathbb{T}^+(u))^{t_0}(\bar{\phi}^{0}_{\l,\l-\eta\hat{\mu}}(-u))^{t_0}
    (\phi^{0}_{\l,\l-\eta\hat{\mu}}(-u))^{t_0}(K^-(u))^{t_0}\rt.\no\\
 &&\qquad\qquad \times \lt.(\tilde{\phi}^{0}_{\l-\eta(\hat{\mu}-\hat{\nu}),\l-\eta\hat{\mu}}(u))^{t_0}
    (\phi^0_{\l-\eta(\hat{\mu}-\hat{\nu}),\l-\eta\hat{\mu}}(u))^{t_0}\rt)\no\\
 &=&\sum_{\mu,\nu}\prod_{k\ne \mu}\frac{\sin(\l_{\mu k}-\eta)}{\sin(\l_{\mu k})}\K(\l|u)_{\mu}^{\nu}\T^+(\l|u)^{\mu}_{\nu}\no\\
 &=&\sum_{\mu}\prod_{k\ne \mu}\frac{\sin(\l_{\mu k}-\eta)}{\sin(\l_{\mu k})}k(\l|u)_{\mu}\T^+(\l|u)^{\mu}_{\mu}.
\eea It was shown that in Ref.\cite{Yan09} the first set of Bethe states given by (\ref{Bethe-state-1})
generated by $\T^{+}(m|u)$ are the eigenstates of our transfer matrix (\ref{trans}) with the eigenvalue
(\ref{Eigenfuction-D-1}) if the parameters $\{v^{(1)}_i\}$ satisfy the Bethe ansatz equation (\ref{BA-D-1}),
while in Ref.\cite{Yan07} the second ones are the eigenstates with the eigenvalue (\ref{Eigenfuction-D-2}) provided that the corresponding
parameters satisfy (\ref{BA-D-2}).

%%%%%%%%%%%%%%%%%%%%%%%%%%%%%%%%%%%%%%%%%%%%%%%%%%%%%%%%%%%%%%%
%                                                             %
%  References                                                 %
%                                                             %
%%%%%%%%%%%%%%%%%%%%%%%%%%%%%%%%%%%%%%%%%%%%%%%%%%%%%%%%%%%%%%%

\end{document}